\RequirePackage{lineno}

\documentclass[aps,prl,twocolumn,superscriptaddress,showpacs,preprintnumbers,amsmath,amssymb]{revtex4-1}

\usepackage{graphicx} 
\usepackage{dcolumn}  
\usepackage{color} %
\RequirePackage{xspace}
\usepackage{relsize}
\usepackage{colortbl}
\usepackage{amsmath} 
\usepackage{ifthen} 
\usepackage{subfigure}

\usepackage[normalem]{ulem}

\usepackage{ifthen} 
\graphicspath{{ps}}
\newboolean{pdflatex}
\setboolean{pdflatex}{true} 
\usepackage{rotating} 

\newboolean{articletitles}
\setboolean{articletitles}{true} 

\newboolean{uprightparticles}
\setboolean{uprightparticles}{false} 

\usepackage{amssymb}
\usepackage{amsfonts}
\usepackage{upgreek} 

\usepackage{hyperref}    
\usepackage[all]{hypcap} 

\begin{document}



\def\lhcb {LHCb\xspace}
\def\ux85 {UX85\xspace}
\def\cern {CERN\xspace}
\def\lhc {LHC\xspace}
\def\atlas {ATLAS\xspace}
\def\cms {CMS\xspace}
\def\babar  {BaBar\xspace}
\def\belle  {Belle\xspace}
\def\aleph  {ALEPH\xspace}
\def\delphi {DELPHI\xspace}
\def\opal   {OPAL\xspace}
\def\lthree {L3\xspace}
\def\lep    {LEP\xspace}
\def\cdf    {CDF\xspace}
\def\dzero  {D\O\xspace}
\def\sld    {SLD\xspace}
\def\cleo   {CLEO\xspace}
\def\uaone  {UA1\xspace}
\def\uatwo  {UA2\xspace}
\def\tevatron {TEVATRON\xspace}


\def\pu     {PU\xspace}
\def\velo   {VELO\xspace}
\def\rich   {RICH\xspace}
\def\richone {RICH1\xspace}
\def\richtwo {RICH2\xspace}
\def\ttracker {TT\xspace}
\def\intr   {IT\xspace}
\def\st     {ST\xspace}
\def\ot     {OT\xspace}
\def\Tone   {T1\xspace}
\def\Ttwo   {T2\xspace}
\def\Tthree {T3\xspace}
\def\Mone   {M1\xspace}
\def\Mtwo   {M2\xspace}
\def\Mthree {M3\xspace}
\def\Mfour  {M4\xspace}
\def\Mfive  {M5\xspace}
\def\ecal   {ECAL\xspace}
\def\spd    {SPD\xspace}
\def\presh  {PS\xspace}
\def\hcal   {HCAL\xspace}
\def\bcm    {BCM\xspace}

\def\ode    {ODE\xspace}
\def\daq    {DAQ\xspace}
\def\tfc    {TFC\xspace}
\def\ecs    {ECS\xspace}
\def\lone   {L0\xspace}
\def\hlt    {HLT\xspace}
\def\hltone {HLT1\xspace}
\def\hlttwo {HLT2\xspace}


\ifthenelse{\boolean{uprightparticles}}%
{\def\Palpha      {\ensuremath{\upalpha}\xspace}
 \def\Pbeta       {\ensuremath{\upbeta}\xspace}
 \def\Pgamma      {\ensuremath{\upgamma}\xspace}                 
 \def\Pdelta      {\ensuremath{\updelta}\xspace}                 
 \def\Pepsilon    {\ensuremath{\upepsilon}\xspace}                 
 \def\Pvarepsilon {\ensuremath{\upvarepsilon}\xspace}                 
 \def\Pzeta       {\ensuremath{\upzeta}\xspace}                 
 \def\Peta        {\ensuremath{\upeta}\xspace}                 
 \def\Ptheta      {\ensuremath{\uptheta}\xspace}                 
 \def\Pvartheta   {\ensuremath{\upvartheta}\xspace}                 
 \def\Piota       {\ensuremath{\upiota}\xspace}                 
 \def\Pkappa      {\ensuremath{\upkappa}\xspace}                 
 \def\Plambda     {\ensuremath{\uplambda}\xspace}                 
 \def\Pmu         {\ensuremath{\upmu}\xspace}                 
 \def\Pnu         {\ensuremath{\upnu}\xspace}                 
 \def\Pxi         {\ensuremath{\upxi}\xspace}                 
 \def\Ppi         {\ensuremath{\uppi}\xspace}                 
 \def\Pvarpi      {\ensuremath{\upvarpi}\xspace}                 
 \def\Prho        {\ensuremath{\uprho}\xspace}                 
 \def\Pvarrho     {\ensuremath{\upvarrho}\xspace}                 
 \def\Ptau        {\ensuremath{\uptau}\xspace}                 
 \def\Pupsilon    {\ensuremath{\upupsilon}\xspace}                 
 \def\Pphi        {\ensuremath{\upphi}\xspace}                 
 \def\Pvarphi     {\ensuremath{\upvarphi}\xspace}                 
 \def\Pchi        {\ensuremath{\upchi}\xspace}                 
 \def\Ppsi        {\ensuremath{\uppsi}\xspace}                 
 \def\Pomega      {\ensuremath{\upomega}\xspace}                 

 \def\PDelta      {\ensuremath{\Delta}\xspace}                 
 \def\PXi      {\ensuremath{\Xi}\xspace}                 
 \def\PLambda      {\ensuremath{\Lambda}\xspace}                 
 \def\PSigma      {\ensuremath{\Sigma}\xspace}                 
 \def\POmega      {\ensuremath{\Omega}\xspace}                 
 \def\PUpsilon      {\ensuremath{\Upsilon}\xspace}                 
 

 \def\PA      {\ensuremath{\mathrm{A}}\xspace}                 
 \def\PB      {\ensuremath{\mathrm{B}}\xspace}                 
 \def\PC      {\ensuremath{\mathrm{C}}\xspace}                 
 \def\PD      {\ensuremath{\mathrm{D}}\xspace}                 
 \def\PE      {\ensuremath{\mathrm{E}}\xspace}                 
 \def\PF      {\ensuremath{\mathrm{F}}\xspace}                 
 \def\PG      {\ensuremath{\mathrm{G}}\xspace}                 
 \def\PH      {\ensuremath{\mathrm{H}}\xspace}                 
 \def\PI      {\ensuremath{\mathrm{I}}\xspace}                 
 \def\PJ      {\ensuremath{\mathrm{J}}\xspace}                 
 \def\PK      {\ensuremath{\mathrm{K}}\xspace}                 
 \def\PL      {\ensuremath{\mathrm{L}}\xspace}                 
 \def\PM      {\ensuremath{\mathrm{M}}\xspace}                 
 \def\PN      {\ensuremath{\mathrm{N}}\xspace}                 
 \def\PO      {\ensuremath{\mathrm{O}}\xspace}                 
 \def\PP      {\ensuremath{\mathrm{P}}\xspace}                 
 \def\PQ      {\ensuremath{\mathrm{Q}}\xspace}                 
 \def\PR      {\ensuremath{\mathrm{R}}\xspace}                 
 \def\PS      {\ensuremath{\mathrm{S}}\xspace}                 
 \def\PT      {\ensuremath{\mathrm{T}}\xspace}                 
 \def\PU      {\ensuremath{\mathrm{U}}\xspace}                 
 \def\PV      {\ensuremath{\mathrm{V}}\xspace}                 
 \def\PW      {\ensuremath{\mathrm{W}}\xspace}                 
 \def\PX      {\ensuremath{\mathrm{X}}\xspace}                 
 \def\PY      {\ensuremath{\mathrm{Y}}\xspace}                 
 \def\PZ      {\ensuremath{\mathrm{Z}}\xspace}                 
 \def\Pa      {\ensuremath{\mathrm{a}}\xspace}                 
 \def\Pb      {\ensuremath{\mathrm{b}}\xspace}                 
 \def\Pc      {\ensuremath{\mathrm{c}}\xspace}                 
 \def\Pd      {\ensuremath{\mathrm{d}}\xspace}                 
 \def\Pe      {\ensuremath{\mathrm{e}}\xspace}                 
 \def\Pf      {\ensuremath{\mathrm{f}}\xspace}                 
 \def\Pg      {\ensuremath{\mathrm{g}}\xspace}                 
 \def\Ph      {\ensuremath{\mathrm{h}}\xspace}                 
 \def\Pi      {\ensuremath{\mathrm{i}}\xspace}                 
 \def\Pj      {\ensuremath{\mathrm{j}}\xspace}                 
 \def\Pk      {\ensuremath{\mathrm{k}}\xspace}                 
 \def\Pl      {\ensuremath{\mathrm{l}}\xspace}                 
 \def\Pm      {\ensuremath{\mathrm{m}}\xspace}                 
 \def\Pn      {\ensuremath{\mathrm{n}}\xspace}                 
 \def\Po      {\ensuremath{\mathrm{o}}\xspace}                 
 \def\Pp      {\ensuremath{\mathrm{p}}\xspace}                 
 \def\Pq      {\ensuremath{\mathrm{q}}\xspace}                 
 \def\Pr      {\ensuremath{\mathrm{r}}\xspace}                 
 \def\Ps      {\ensuremath{\mathrm{s}}\xspace}                 
 \def\Pt      {\ensuremath{\mathrm{t}}\xspace}                 
 \def\Pu      {\ensuremath{\mathrm{u}}\xspace}                 
 \def\Pv      {\ensuremath{\mathrm{v}}\xspace}                 
 \def\Pw      {\ensuremath{\mathrm{w}}\xspace}                 
 \def\Px      {\ensuremath{\mathrm{x}}\xspace}                 
 \def\Py      {\ensuremath{\mathrm{y}}\xspace}                 
 \def\Pz      {\ensuremath{\mathrm{z}}\xspace}                 
}
{\def\Palpha      {\ensuremath{\alpha}\xspace}
 \def\Pbeta       {\ensuremath{\beta}\xspace}
 \def\Pgamma      {\ensuremath{\gamma}\xspace}                 
 \def\Pdelta      {\ensuremath{\delta}\xspace}                 
 \def\Pepsilon    {\ensuremath{\epsilon}\xspace}                 
 \def\Pvarepsilon {\ensuremath{\varepsilon}\xspace}                 
 \def\Pzeta       {\ensuremath{\zeta}\xspace}                 
 \def\Peta        {\ensuremath{\eta}\xspace}                 
 \def\Ptheta      {\ensuremath{\theta}\xspace}                 
 \def\Pvartheta   {\ensuremath{\vartheta}\xspace}                 
 \def\Piota       {\ensuremath{\iota}\xspace}                 
 \def\Pkappa      {\ensuremath{\kappa}\xspace}                 
 \def\Plambda     {\ensuremath{\lambda}\xspace}                 
 \def\Pmu         {\ensuremath{\mu}\xspace}                 
 \def\Pnu         {\ensuremath{\nu}\xspace}                 
 \def\Pxi         {\ensuremath{\xi}\xspace}                 
 \def\Ppi         {\ensuremath{\pi}\xspace}                 
 \def\Pvarpi      {\ensuremath{\varpi}\xspace}                 
 \def\Prho        {\ensuremath{\rho}\xspace}                 
 \def\Pvarrho     {\ensuremath{\varrho}\xspace}                 
 \def\Ptau        {\ensuremath{\tau}\xspace}                 
 \def\Pupsilon    {\ensuremath{\upsilon}\xspace}                 
 \def\Pphi        {\ensuremath{\phi}\xspace}                 
 \def\Pvarphi     {\ensuremath{\varphi}\xspace}                 
 \def\Pchi        {\ensuremath{\chi}\xspace}                 
 \def\Ppsi        {\ensuremath{\psi}\xspace}                 
 \def\Pomega      {\ensuremath{\omega}\xspace}                 
 \mathchardef\PDelta="7101
 \mathchardef\PXi="7104
 \mathchardef\PLambda="7103
 \mathchardef\PSigma="7106
 \mathchardef\POmega="710A
 \mathchardef\PUpsilon="7107
 \def\PA      {\ensuremath{A}\xspace}                 
 \def\PB      {\ensuremath{B}\xspace}                 
 \def\PC      {\ensuremath{C}\xspace}                 
 \def\PD      {\ensuremath{D}\xspace}                 
 \def\PE      {\ensuremath{E}\xspace}                 
 \def\PF      {\ensuremath{F}\xspace}                 
 \def\PG      {\ensuremath{G}\xspace}                 
 \def\PH      {\ensuremath{H}\xspace}                 
 \def\PI      {\ensuremath{I}\xspace}                 
 \def\PJ      {\ensuremath{J}\xspace}                 
 \def\PK      {\ensuremath{K}\xspace}                 
 \def\PL      {\ensuremath{L}\xspace}                 
 \def\PM      {\ensuremath{M}\xspace}                 
 \def\PN      {\ensuremath{N}\xspace}                 
 \def\PO      {\ensuremath{O}\xspace}                 
 \def\PP      {\ensuremath{P}\xspace}                 
 \def\PQ      {\ensuremath{Q}\xspace}                 
 \def\PR      {\ensuremath{R}\xspace}                 
 \def\PS      {\ensuremath{S}\xspace}                 
 \def\PT      {\ensuremath{T}\xspace}                 
 \def\PU      {\ensuremath{U}\xspace}                 
 \def\PV      {\ensuremath{V}\xspace}                 
 \def\PW      {\ensuremath{W}\xspace}                 
 \def\PX      {\ensuremath{X}\xspace}                 
 \def\PY      {\ensuremath{Y}\xspace}                 
 \def\PZ      {\ensuremath{Z}\xspace}                 
 \def\Pa      {\ensuremath{a}\xspace}                 
 \def\Pb      {\ensuremath{b}\xspace}                 
 \def\Pc      {\ensuremath{c}\xspace}                 
 \def\Pd      {\ensuremath{d}\xspace}                 
 \def\Pe      {\ensuremath{e}\xspace}                 
 \def\Pf      {\ensuremath{f}\xspace}                 
 \def\Pg      {\ensuremath{g}\xspace}                 
 \def\Ph      {\ensuremath{h}\xspace}                 
 \def\Pi      {\ensuremath{i}\xspace}                 
 \def\Pj      {\ensuremath{j}\xspace}                 
 \def\Pk      {\ensuremath{k}\xspace}                 
 \def\Pl      {\ensuremath{l}\xspace}                 
 \def\Pm      {\ensuremath{m}\xspace}                 
 \def\Pn      {\ensuremath{n}\xspace}                 
 \def\Po      {\ensuremath{o}\xspace}                 
 \def\Pp      {\ensuremath{p}\xspace}                 
 \def\Pq      {\ensuremath{q}\xspace}                 
 \def\Pr      {\ensuremath{r}\xspace}                 
 \def\Ps      {\ensuremath{s}\xspace}                 
 \def\Pt      {\ensuremath{t}\xspace}                 
 \def\Pu      {\ensuremath{u}\xspace}                 
 \def\Pv      {\ensuremath{v}\xspace}                 
 \def\Pw      {\ensuremath{w}\xspace}                 
 \def\Px      {\ensuremath{x}\xspace}                 
 \def\Py      {\ensuremath{y}\xspace}                 
 \def\Pz      {\ensuremath{z}\xspace}                 
}



\let\emi\en
\def\electron   {\ensuremath{\Pe}\xspace}
\def\en         {\ensuremath{\Pe^-}\xspace}   
\def\ep         {\ensuremath{\Pe^+}\xspace}
\def\epm        {\ensuremath{\Pe^\pm}\xspace} 
\def\epem       {\ensuremath{\Pe^+\Pe^-}\xspace}
\def\ee         {\ensuremath{\Pe^-\Pe^-}\xspace}

\def\mmu        {\ensuremath{\Pmu}\xspace}
\def\mup        {\ensuremath{\Pmu^+}\xspace}
\def\mun        {\ensuremath{\Pmu^-}\xspace} 
\def\mumu       {\ensuremath{\Pmu^+\Pmu^-}\xspace}
\def\mtau       {\ensuremath{\Ptau}\xspace}

\def\taup       {\ensuremath{\Ptau^+}\xspace}
\def\taum       {\ensuremath{\Ptau^-}\xspace}
\def\tautau     {\ensuremath{\Ptau^+\Ptau^-}\xspace}

\def\ellm       {\ensuremath{\ell^-}\xspace}
\def\ellp       {\ensuremath{\ell^+}\xspace}
\def\ellell     {\ensuremath{\ell^+ \ell^-}\xspace}

\def\neu        {\ensuremath{\Pnu}\xspace}
\def\neub       {\ensuremath{\overline{\Pnu}}\xspace}
\def\nuenueb    {\ensuremath{\neu\neub}\xspace}
\def\neue       {\ensuremath{\neu_e}\xspace}
\def\neueb      {\ensuremath{\neub_e}\xspace}
\def\neueneueb  {\ensuremath{\neue\neueb}\xspace}
\def\neum       {\ensuremath{\neu_\mu}\xspace}
\def\neumb      {\ensuremath{\neub_\mu}\xspace}
\def\neumneumb  {\ensuremath{\neum\neumb}\xspace}
\def\neut       {\ensuremath{\neu_\tau}\xspace}
\def\neutb      {\ensuremath{\neub_\tau}\xspace}
\def\neutneutb  {\ensuremath{\neut\neutb}\xspace}
\def\neul       {\ensuremath{\neu_\ell}\xspace}
\def\neulb      {\ensuremath{\neub_\ell}\xspace}
\def\neulneulb  {\ensuremath{\neul\neulb}\xspace}


\def\g      {\ensuremath{\Pgamma}\xspace}
\def\H      {\ensuremath{\PH^0}\xspace}
\def\Hp     {\ensuremath{\PH^+}\xspace}
\def\Hm     {\ensuremath{\PH^-}\xspace}
\def\Hpm    {\ensuremath{\PH^\pm}\xspace}
\def\W      {\ensuremath{\PW}\xspace}
\def\Wp     {\ensuremath{\PW^+}\xspace}
\def\Wm     {\ensuremath{\PW^-}\xspace}
\def\Wpm    {\ensuremath{\PW^\pm}\xspace}
\def\Z      {\ensuremath{\PZ^0}\xspace}


\def\quark     {\ensuremath{\Pq}\xspace}
\def\quarkbar  {\ensuremath{\overline \quark}\xspace}
\def\qqbar     {\ensuremath{\quark\quarkbar}\xspace}
\def\uquark    {\ensuremath{\Pu}\xspace}
\def\uquarkbar {\ensuremath{\overline \uquark}\xspace}
\def\uubar     {\ensuremath{\uquark\uquarkbar}\xspace}
\def\dquark    {\ensuremath{\Pd}\xspace}
\def\dquarkbar {\ensuremath{\overline \dquark}\xspace}
\def\ddbar     {\ensuremath{\dquark\dquarkbar}\xspace}
\def\squark    {\ensuremath{\Ps}\xspace}
\def\squarkbar {\ensuremath{\overline \squark}\xspace}
\def\ssbar     {\ensuremath{\squark\squarkbar}\xspace}
\def\cquark    {\ensuremath{\Pc}\xspace}
\def\cquarkbar {\ensuremath{\overline \cquark}\xspace}
\def\ccbar     {\ensuremath{\cquark\cquarkbar}\xspace}
\def\bquark    {\ensuremath{\Pb}\xspace}
\def\bquarkbar {\ensuremath{\overline \bquark}\xspace}
\def\bbbar     {\ensuremath{\bquark\bquarkbar}\xspace}
\def\tquark    {\ensuremath{\Pt}\xspace}
\def\tquarkbar {\ensuremath{\overline \tquark}\xspace}
\def\ttbar     {\ensuremath{\tquark\tquarkbar}\xspace}


\def\pion  {\ensuremath{\Ppi}\xspace}
\def\piz   {\ensuremath{\pion^0}\xspace}
\def\pizs  {\ensuremath{\pion^0\mbox\,\rm{s}}\xspace}
\def\ppz   {\ensuremath{\pion^0\pion^0}\xspace}
\def\pip   {\ensuremath{\pion^+}\xspace}
\def\pim   {\ensuremath{\pion^-}\xspace}
\def\pipi  {\ensuremath{\pion^+\pion^-}\xspace}
\def\pipm  {\ensuremath{\pion^\pm}\xspace}
\def\pimp  {\ensuremath{\pion^\mp}\xspace}

\def\kaon  {\ensuremath{\PK}\xspace}
  \def\Kbar  {\kern 0.2em\overline{\kern -0.2em \PK}{}\xspace}
\def\Kb    {\ensuremath{\Kbar}\xspace}
\def\Kz    {\ensuremath{\kaon^0}\xspace}
\def\Kzb   {\ensuremath{\Kbar^0}\xspace}
\def\KzKzb {\ensuremath{\Kz \kern -0.16em \Kzb}\xspace}
\def\Kp    {\ensuremath{\kaon^+}\xspace}
\def\Km    {\ensuremath{\kaon^-}\xspace}
\def\Kpm   {\ensuremath{\kaon^\pm}\xspace}
\def\Kmp   {\ensuremath{\kaon^\mp}\xspace}
\def\KpKm  {\ensuremath{\Kp \kern -0.16em \Km}\xspace}
\def\KS    {\ensuremath{\kaon^0_{\rm\scriptscriptstyle S}}\xspace} 
\def\KSb    {\ensuremath{\Kbar^0_{\rm\scriptscriptstyle S}}\xspace} 
\def\KL    {\ensuremath{\kaon^0_{\rm\scriptscriptstyle L}}\xspace} 
\def\Kstarz  {\ensuremath{\kaon^{*0}}\xspace}
\def\Kstarzb {\ensuremath{\Kbar^{*0}}\xspace}
\def\Kstar   {\ensuremath{\kaon^*}\xspace}
\def\Kstarb  {\ensuremath{\Kbar^*}\xspace}
\def\Kstarp  {\ensuremath{\kaon^{*+}}\xspace}
\def\Kstarm  {\ensuremath{\kaon^{*-}}\xspace}
\def\Kstarpm {\ensuremath{\kaon^{*\pm}}\xspace}
\def\Kstarmp {\ensuremath{\kaon^{*\mp}}\xspace}

\newcommand{\etapr}{\ensuremath{\Peta^{\prime}}\xspace}


  \def\Dbar    {\kern 0.2em\overline{\kern -0.2em \PD}{}\xspace}
\def\D       {\ensuremath{\PD}\xspace}
\def\Db      {\ensuremath{\Dbar}\xspace}
\def\Dz      {\ensuremath{\D^0}\xspace}
\def\Dzb     {\ensuremath{\Dbar^0}\xspace}
\def\DzDzb   {\ensuremath{\Dz {\kern -0.16em \Dzb}}\xspace}
\def\Dp      {\ensuremath{\D^+}\xspace}
\def\Dm      {\ensuremath{\D^-}\xspace}
\def\Dpm     {\ensuremath{\D^\pm}\xspace}
\def\Dmp     {\ensuremath{\D^\mp}\xspace}
\def\DpDm    {\ensuremath{\Dp {\kern -0.16em \Dm}}\xspace}
\def\Dstar   {\ensuremath{\D^*}\xspace}
\def\Dstarb  {\ensuremath{\Dbar^*}\xspace}
\def\Dstarz  {\ensuremath{\D^{*0}}\xspace}
\def\Dstarzb {\ensuremath{\Dbar^{*0}}\xspace}
\def\Dstarp  {\ensuremath{\D^{*+}}\xspace}
\def\Dstarm  {\ensuremath{\D^{*-}}\xspace}
\def\Dstarpm {\ensuremath{\D^{*\pm}}\xspace}
\def\Dstarmp {\ensuremath{\D^{*\mp}}\xspace}
\def\Ds      {\ensuremath{\D^+_\squark}\xspace}
\def\Dsp     {\ensuremath{\D^+_\squark}\xspace}
\def\Dsm     {\ensuremath{\D^-_\squark}\xspace}
\def\Dspm    {\ensuremath{\D^{\pm}_\squark}\xspace}
\def\Dss     {\ensuremath{\D^{*+}_\squark}\xspace}
\def\Dssp    {\ensuremath{\D^{*+}_\squark}\xspace}
\def\Dssm    {\ensuremath{\D^{*-}_\squark}\xspace}
\def\Dsspm   {\ensuremath{\D^{*\pm}_\squark}\xspace}

\def\B       {\ensuremath{\PB}\xspace}
  \def\Bbar    {\kern 0.18em\overline{\kern -0.18em \PB}{}\xspace}
\def\Bb      {\ensuremath{\Bbar}\xspace}
\def\BBbar   {\ensuremath{\B\Bbar}\xspace} 
\def\Bz      {\ensuremath{\B^0}\xspace}
\def\Bzb     {\ensuremath{\Bbar^0}\xspace}
\def\Bu      {\ensuremath{\B^+}\xspace}
\def\Bub     {\ensuremath{\B^-}\xspace}
\def\Bp      {\ensuremath{\Bu}\xspace}
\def\Bm      {\ensuremath{\Bub}\xspace}
\def\Bpm     {\ensuremath{\B^\pm}\xspace}
\def\Bmp     {\ensuremath{\B^\mp}\xspace}
\def\Bd      {\ensuremath{\B^0}\xspace}
\def\Bs      {\ensuremath{\B^0_\squark}\xspace}
\def\Bsb     {\ensuremath{\Bbar^0_\squark}\xspace}
\def\Bstar   {\ensuremath{B_s^*}\xspace}
\def\Bstarb  {\ensuremath{\Bb_s^*}\xspace}
\def\Bdb     {\ensuremath{\Bbar^0}\xspace}
\def\Bc      {\ensuremath{\B_\cquark^+}\xspace}
\def\Bcp     {\ensuremath{\B_\cquark^+}\xspace}
\def\Bcm     {\ensuremath{\B_\cquark^-}\xspace}
\def\Bcpm    {\ensuremath{\B_\cquark^\pm}\xspace}


\def\jpsi     {\ensuremath{{\PJ\mskip -3mu/\mskip -2mu\Ppsi\mskip 2mu}}\xspace}
\def\psitwos  {\ensuremath{\Ppsi{(2S)}}\xspace}
\def\psiprpr  {\ensuremath{\Ppsi(3770)}\xspace}
\def\etac     {\ensuremath{\Peta_\cquark}\xspace}
\def\chiczero {\ensuremath{\Pchi_{\cquark 0}}\xspace}
\def\chicone  {\ensuremath{\Pchi_{\cquark 1}}\xspace}
\def\chictwo  {\ensuremath{\Pchi_{\cquark 2}}\xspace}
  \def\Y#1S{\ensuremath{\PUpsilon{(#1S)}}\xspace}
\def\OneS  {\Y1S}
\def\TwoS  {\Y2S}
\def\ThreeS{\Y3S}
\def\FourS {\Y4S}
\def\FiveS {\Y5S}

\def\chic  {\ensuremath{\Pchi_{c}}\xspace}


\def\proton      {\ensuremath{\Pp}\xspace}
\def\antiproton  {\ensuremath{\overline \proton}\xspace}
\def\neutron     {\ensuremath{\Pn}\xspace}
\def\antineutron {\ensuremath{\overline \neutron}\xspace}

\def\Deltares {\ensuremath{\PDelta}\xspace}
\def\Deltaresbar{\ensuremath{\overline \Deltares}\xspace}
\def\Xires {\ensuremath{\PXi}\xspace}
\def\Xiresbar{\ensuremath{\overline \Xires}\xspace}
\def\L {\ensuremath{\PLambda}\xspace}
\def\Lbar {\ensuremath{\kern 0.1em\overline{\kern -0.1em\Lambda\kern -0.05em}\kern 0.05em{}}\xspace}
\def\Lambdares {\ensuremath{\PLambda}\xspace}
\def\Lambdaresbar{\ensuremath{\Lbar}\xspace}
\def\Sigmares {\ensuremath{\PSigma}\xspace}
\def\Sigmaresbar{\ensuremath{\overline \Sigmares}\xspace}
\def\Omegares {\ensuremath{\POmega}\xspace}
\def\Omegaresbar{\ensuremath{\overline \Omegares}\xspace}


\def\Lb      {\ensuremath{\L^0_\bquark}\xspace}
\def\Lbbar   {\ensuremath{\Lbar^0_\bquark}\xspace}
\def\Lc      {\ensuremath{\L^+_\cquark}\xspace}
\def\Lcbar   {\ensuremath{\Lbar^-_\cquark}\xspace}


\def\BF         {{\ensuremath{\cal B}\xspace}}
\def\BRvis      {{\ensuremath{\BR_{\rm{vis}}}}}
\def\BR         {\BF}
\newcommand{\decay}[2]{\ensuremath{#1\!\to #2}\xspace}         
\def\ra                 {\ensuremath{\rightarrow}\xspace}
\def\to                 {\ensuremath{\rightarrow}\xspace}

\newcommand{\tauBs}{\ensuremath{\tau_{\Bs}}\xspace}
\newcommand{\tauBd}{\ensuremath{\tau_{\Bd}}\xspace}
\newcommand{\tauBz}{\ensuremath{\tau_{\Bz}}\xspace}
\newcommand{\tauBu}{\ensuremath{\tau_{\Bp}}\xspace}
\newcommand{\tauDp}{\ensuremath{\tau_{\Dp}}\xspace}
\newcommand{\tauDz}{\ensuremath{\tau_{\Dz}}\xspace}
\newcommand{\tauL}{\ensuremath{\tau_{\rm L}}\xspace}
\newcommand{\tauH}{\ensuremath{\tau_{\rm H}}\xspace}

\newcommand{\mBd}{\ensuremath{m_{\Bd}}\xspace}
\newcommand{\mBp}{\ensuremath{m_{\Bp}}\xspace}
\newcommand{\mBs}{\ensuremath{m_{\Bs}}\xspace}
\newcommand{\mBc}{\ensuremath{m_{\Bc}}\xspace}
\newcommand{\mLb}{\ensuremath{m_{\Lb}}\xspace}

\def\grpsuthree {\ensuremath{\mathrm{SU}(3)}\xspace}
\def\grpsutw    {\ensuremath{\mathrm{SU}(2)}\xspace}
\def\grpuone    {\ensuremath{\mathrm{U}(1)}\xspace}

\def\ssqtw {\ensuremath{\sin^{2}\!\theta_{\mathrm{W}}}\xspace}
\def\csqtw {\ensuremath{\cos^{2}\!\theta_{\mathrm{W}}}\xspace}
\def\stw   {\ensuremath{\sin\theta_{\mathrm{W}}}\xspace}
\def\ctw   {\ensuremath{\cos\theta_{\mathrm{W}}}\xspace}
\def\ssqtwef {\ensuremath{{\sin}^{2}\theta_{\mathrm{W}}^{\mathrm{eff}}}\xspace}
\def\csqtwef {\ensuremath{{\cos}^{2}\theta_{\mathrm{W}}^{\mathrm{eff}}}\xspace}
\def\stwef {\ensuremath{\sin\theta_{\mathrm{W}}^{\mathrm{eff}}}\xspace}
\def\ctwef {\ensuremath{\cos\theta_{\mathrm{W}}^{\mathrm{eff}}}\xspace}
\def\gv    {\ensuremath{g_{\mbox{\tiny V}}}\xspace}
\def\ga    {\ensuremath{g_{\mbox{\tiny A}}}\xspace}

\def\order   {\ensuremath{\mathcal{O}}\xspace}
\def\ordalph {\ensuremath{\mathcal{O}(\alpha)}\xspace}
\def\ordalsq {\ensuremath{\mathcal{O}(\alpha^{2})}\xspace}
\def\ordalcb {\ensuremath{\mathcal{O}(\alpha^{3})}\xspace}

\newcommand{\as}{\ensuremath{\alpha_{\scriptscriptstyle S}}\xspace}
\newcommand{\MSb}{\ensuremath{\overline{\mathrm{MS}}}\xspace}
\newcommand{\lqcd}{\ensuremath{\Lambda_{\mathrm{QCD}}}\xspace}
\def\qsq       {\ensuremath{q^2}\xspace}


\def\eps   {\ensuremath{\varepsilon}\xspace}
\def\epsK  {\ensuremath{\varepsilon_K}\xspace}
\def\epsB  {\ensuremath{\varepsilon_B}\xspace}
\def\epsp  {\ensuremath{\varepsilon^\prime_K}\xspace}

\def\CP                {\ensuremath{C\!P}\xspace}
\def\CPT               {\ensuremath{C\!PT}\xspace}

\def\rhobar {\ensuremath{\overline \rho}\xspace}
\def\etabar {\ensuremath{\overline \eta}\xspace}

\def\Vud  {\ensuremath{|V_{\uquark\dquark}|}\xspace}
\def\Vcd  {\ensuremath{|V_{\cquark\dquark}|}\xspace}
\def\Vtd  {\ensuremath{|V_{\tquark\dquark}|}\xspace}
\def\Vus  {\ensuremath{|V_{\uquark\squark}|}\xspace}
\def\Vcs  {\ensuremath{|V_{\cquark\squark}|}\xspace}
\def\Vts  {\ensuremath{|V_{\tquark\squark}|}\xspace}
\def\Vub  {\ensuremath{|V_{\uquark\bquark}|}\xspace}
\def\Vcb  {\ensuremath{|V_{\cquark\bquark}|}\xspace}
\def\Vtb  {\ensuremath{|V_{\tquark\bquark}|}\xspace}


\newcommand{\dm}{\ensuremath{\Delta m}\xspace}
\newcommand{\dms}{\ensuremath{\Delta m_{\squark}}\xspace}
\newcommand{\dmd}{\ensuremath{\Delta m_{\dquark}}\xspace}
\newcommand{\DG}{\ensuremath{\Delta\Gamma}\xspace}
\newcommand{\DGs}{\ensuremath{\Delta\Gamma_{\squark}}\xspace}
\newcommand{\DGd}{\ensuremath{\Delta\Gamma_{\dquark}}\xspace}
\newcommand{\Gs}{\ensuremath{\Gamma_{\squark}}\xspace}
\newcommand{\Gd}{\ensuremath{\Gamma_{\dquark}}\xspace}

\newcommand{\MBq}{\ensuremath{M_{\B_\quark}}\xspace}
\newcommand{\DGq}{\ensuremath{\Delta\Gamma_{\quark}}\xspace}
\newcommand{\Gq}{\ensuremath{\Gamma_{\quark}}\xspace}
\newcommand{\dmq}{\ensuremath{\Delta m_{\quark}}\xspace}
\newcommand{\GL}{\ensuremath{\Gamma_{\rm L}}\xspace}
\newcommand{\GH}{\ensuremath{\Gamma_{\rm H}}\xspace}

\newcommand{\DGsGs}{\ensuremath{\Delta\Gamma_{\squark}/\Gamma_{\squark}}\xspace}
\newcommand{\Delm}{\mbox{$\Delta m $}\xspace}
\newcommand{\ACP}{\ensuremath{{\cal A}^{\CP}}\xspace}
\newcommand{\Adir}{\ensuremath{{\cal A}^{\rm dir}}\xspace}
\newcommand{\Amix}{\ensuremath{{\cal A}^{\rm mix}}\xspace}
\newcommand{\ADelta}{\ensuremath{{\cal A}^\Delta}\xspace}
\newcommand{\phid}{\ensuremath{\phi_{\dquark}}\xspace}
\newcommand{\sinphid}{\ensuremath{\sin\!\phid}\xspace}
\newcommand{\phis}{\ensuremath{\phi_{\squark}}\xspace}
\newcommand{\betas}{\ensuremath{\beta_{\squark}}\xspace}
\newcommand{\sbetas}{\ensuremath{\sigma(\beta_{\squark})}\xspace}
\newcommand{\stbetas}{\ensuremath{\sigma(2\beta_{\squark})}\xspace}
\newcommand{\stphis}{\ensuremath{\sigma(\phi_{\squark})}\xspace}
\newcommand{\sinphis}{\ensuremath{\sin\!\phis}\xspace}

\newcommand{\edet}{{\ensuremath{\varepsilon_{\rm det}}}\xspace}
\newcommand{\erec}{{\ensuremath{\varepsilon_{\rm rec/det}}}\xspace}
\newcommand{\esel}{{\ensuremath{\varepsilon_{\rm sel/rec}}}\xspace}
\newcommand{\etrg}{{\ensuremath{\varepsilon_{\rm trg/sel}}}\xspace}
\newcommand{\etot}{{\ensuremath{\varepsilon_{\rm tot}}}\xspace}

\newcommand{\mistag}{\ensuremath{\omega}\xspace}
\newcommand{\wcomb}{\ensuremath{\omega^{\rm comb}}\xspace}
\newcommand{\etag}{{\ensuremath{\varepsilon_{\rm tag}}}\xspace}
\newcommand{\etagcomb}{{\ensuremath{\varepsilon_{\rm tag}^{\rm comb}}}\xspace}
\newcommand{\effeff}{\ensuremath{\varepsilon_{\rm eff}}\xspace}
\newcommand{\effeffcomb}{\ensuremath{\varepsilon_{\rm eff}^{\rm comb}}\xspace}
\newcommand{\efftag}{{\ensuremath{\etag(1-2\omega)^2}}\xspace}
\newcommand{\effD}{{\ensuremath{\etag D^2}}\xspace}

\newcommand{\etagprompt}{{\ensuremath{\varepsilon_{\rm tag}^{\rm Pr}}}\xspace}
\newcommand{\etagLL}{{\ensuremath{\varepsilon_{\rm tag}^{\rm LL}}}\xspace}


\def\BdToKstmm    {\decay{\Bd}{\Kstarz\mup\mun}}
\def\BdbToKstmm   {\decay{\Bdb}{\Kstarzb\mup\mun}}

\def\BsToJPsiPhi  {\decay{\Bs}{\jpsi\phi}}
\def\BdToJPsiKst  {\decay{\Bd}{\jpsi\Kstarz}}
\def\BdbToJPsiKst {\decay{\Bdb}{\jpsi\Kstarzb}}

\def\BsPhiGam     {\decay{\Bs}{\phi \g}}
\def\BdKstGam     {\decay{\Bd}{\Kstarz \g}}

\def\BTohh        {\decay{\B}{\Ph^+ \Ph'^-}}
\def\BdTopipi     {\decay{\Bd}{\pip\pim}}
\def\BdToKpi      {\decay{\Bd}{\Kp\pim}}
\def\BsToKK       {\decay{\Bs}{\Kp\Km}}
\def\BsTopiK      {\decay{\Bs}{\pip\Km}}

\def\BdKstee  {\decay{\Bd}{\Kstarz\epem}}
\def\BdbKstee {\decay{\Bdb}{\Kstarzb\epem}}
\def\bsll     {\decay{\bquark}{\squark \ell^+ \ell^-}}
\def\AFB      {\ensuremath{A_{\mathrm{FB}}}\xspace}
\def\FL       {\ensuremath{F_{\mathrm{L}}}\xspace}
\def\AT#1     {\ensuremath{A_{\mathrm{T}}^{#1}}\xspace}           
\def\btosgam  {\decay{\bquark}{\squark \g}}
\def\btodgam  {\decay{\bquark}{\dquark \g}}
\def\Bsmm     {\decay{\Bs}{\mup\mun}}
\def\Bdmm     {\decay{\Bd}{\mup\mun}}
\def\ctl       {\ensuremath{\cos{\theta_l}}\xspace}
\def\ctk       {\ensuremath{\cos{\theta_K}}\xspace}

\def\C#1      {\ensuremath{\mathcal{C}_{#1}}\xspace}                       
\def\Cp#1     {\ensuremath{\mathcal{C}_{#1}^{'}}\xspace}                    
\def\Ceff#1   {\ensuremath{\mathcal{C}_{#1}^{\mathrm{(eff)}}}\xspace}        
\def\Cpeff#1  {\ensuremath{\mathcal{C}_{#1}^{'\mathrm{(eff)}}}\xspace}       
\def\Ope#1    {\ensuremath{\mathcal{O}_{#1}}\xspace}                       
\def\Opep#1   {\ensuremath{\mathcal{O}_{#1}^{'}}\xspace}                    


\def\xprime     {\ensuremath{x^{\prime}}\xspace}
\def\yprime     {\ensuremath{y^{\prime}}\xspace}
\def\ycp        {\ensuremath{y_{\CP}}\xspace}
\def\agamma     {\ensuremath{A_{\Gamma}}\xspace}
\def\kpi        {\ensuremath{\PK\Ppi}\xspace}
\def\kk         {\ensuremath{\PK\PK}\xspace}
\def\dkpi       {\decay{\PD}{\PK\Ppi}}
\def\dkk        {\decay{\PD}{\PK\PK}}
\def\dkpicf     {\decay{\Dz}{\Km\pip}}

\newcommand{\bra}[1]{\ensuremath{\langle #1|}}             
\newcommand{\ket}[1]{\ensuremath{|#1\rangle}}              
\newcommand{\braket}[2]{\ensuremath{\langle #1|#2\rangle}} 

\newcommand{\unit}[1]{\ensuremath{\rm\,#1}\xspace}          

\newcommand{\tev}{\ensuremath{\mathrm{\,Te\kern -0.1em V}}\xspace}
\newcommand{\gev}{\ensuremath{\mathrm{\,Ge\kern -0.1em V}}\xspace}
\newcommand{\mev}{\ensuremath{\mathrm{\,Me\kern -0.1em V}}\xspace}
\newcommand{\kev}{\ensuremath{\mathrm{\,ke\kern -0.1em V}}\xspace}
\newcommand{\ev}{\ensuremath{\mathrm{\,e\kern -0.1em V}}\xspace}
\newcommand{\gevc}{\ensuremath{{\mathrm{\,Ge\kern -0.1em V\!/}c}}\xspace}
\newcommand{\mevc}{\ensuremath{{\mathrm{\,Me\kern -0.1em V\!/}c}}\xspace}
\newcommand{\gevcc}{\ensuremath{{\mathrm{\,Ge\kern -0.1em V\!/}c^2}}\xspace}
\newcommand{\gevgevcccc}{\ensuremath{{\mathrm{\,Ge\kern -0.1em V^2\!/}c^4}}\xspace}
\newcommand{\mevcc}{\ensuremath{{\mathrm{\,Me\kern -0.1em V\!/}c^2}}\xspace}

\def\km   {\ensuremath{\rm \,km}\xspace}
\def\m    {\ensuremath{\rm \,m}\xspace}
\def\cm   {\ensuremath{\rm \,cm}\xspace}
\def\cma  {\ensuremath{{\rm \,cm}^2}\xspace}
\def\mm   {\ensuremath{\rm \,mm}\xspace}
\def\mma  {\ensuremath{{\rm \,mm}^2}\xspace}
\def\mum  {\ensuremath{\,\upmu\rm m}\xspace}
\def\muma {\ensuremath{\,\upmu\rm m^2}\xspace}
\def\nm   {\ensuremath{\rm \,nm}\xspace}
\def\fm   {\ensuremath{\rm \,fm}\xspace}
\def\barn{\ensuremath{\rm \,b}\xspace}
\def\barnhyph{\ensuremath{\rm -b}\xspace}
\def\mbarn{\ensuremath{\rm \,mb}\xspace}
\def\mub{\ensuremath{\rm \,\upmu b}\xspace}
\def\mbarnhyph{\ensuremath{\rm -mb}\xspace}
\def\nb {\ensuremath{\rm \,nb}\xspace}
\def\invnb {\ensuremath{\mbox{\,nb}^{-1}}\xspace}
\def\pb {\ensuremath{\rm \,pb}\xspace}
\def\invpb {\ensuremath{\mbox{\,pb}^{-1}}\xspace}
\def\fb   {\ensuremath{\mbox{\,fb}}\xspace}
\def\invfb   {\ensuremath{\mbox{\,fb}^{-1}}\xspace}

\def\sec  {\ensuremath{\rm {\,s}}\xspace}
\def\ms   {\ensuremath{{\rm \,ms}}\xspace}
\def\mus  {\ensuremath{\,\upmu{\rm s}}\xspace}
\def\ns   {\ensuremath{{\rm \,ns}}\xspace}
\def\ps   {\ensuremath{{\rm \,ps}}\xspace}
\def\fs   {\ensuremath{\rm \,fs}\xspace}

\def\mhz  {\ensuremath{{\rm \,MHz}}\xspace}
\def\khz  {\ensuremath{{\rm \,kHz}}\xspace}
\def\hz   {\ensuremath{{\rm \,Hz}}\xspace}

\def\invps{\ensuremath{{\rm \,ps^{-1}}}\xspace}

\def\yr   {\ensuremath{\rm \,yr}\xspace}
\def\hr   {\ensuremath{\rm \,hr}\xspace}

\def\degc {\ensuremath{^\circ}{C}\xspace}
\def\degk {\ensuremath {\rm K}\xspace}

\def\Xrad {\ensuremath{X_0}\xspace}
\def\NIL{\ensuremath{\lambda_{int}}\xspace}
\def\mip {MIP\xspace}
\def\neutroneq {\ensuremath{\rm \,n_{eq}}\xspace}
\def\neqcmcm {\ensuremath{\rm \,n_{eq} / cm^2}\xspace}
\def\kRad {\ensuremath{\rm \,kRad}\xspace}
\def\MRad {\ensuremath{\rm \,MRad}\xspace}
\def\ci {\ensuremath{\rm \,Ci}\xspace}
\def\mci {\ensuremath{\rm \,mCi}\xspace}

\def\sx    {\ensuremath{\sigma_x}\xspace}    
\def\sy    {\ensuremath{\sigma_y}\xspace}   
\def\sz    {\ensuremath{\sigma_z}\xspace}    

\newcommand{\stat}{\ensuremath{\mathrm{(stat)}}\xspace}
\newcommand{\syst}{\ensuremath{\mathrm{(syst)}}\xspace}


\def\order{{\ensuremath{\cal O}}\xspace}
\newcommand{\chisq}{\ensuremath{\chi^2}\xspace}

\def\deriv {\ensuremath{\mathrm{d}}}

\def\gsim{{~\raise.15em\hbox{$>$}\kern-.85em
          \lower.35em\hbox{$\sim$}~}\xspace}
\def\lsim{{~\raise.15em\hbox{$<$}\kern-.85em
          \lower.35em\hbox{$\sim$}~}\xspace}

\newcommand{\mean}[1]{\ensuremath{\left\langle #1 \right\rangle}} 
\newcommand{\abs}[1]{\ensuremath{\left\|#1\right\|}} 
\newcommand{\Real}{\ensuremath{\mathcal{R}e}\xspace}
\newcommand{\Imag}{\ensuremath{\mathcal{I}m}\xspace}

\def\PDF {PDF\xspace}

\def\Ebeam {\ensuremath{E_{\mbox{\tiny BEAM}}}\xspace}
\def\sqs   {\ensuremath{\protect\sqrt{s}}\xspace}

\def\ptot       {\mbox{$p$}\xspace}
\def\pt         {\mbox{$p_{\rm T}$}\xspace}
\def\et         {\mbox{$E_{\rm T}$}\xspace}
\def\dpp        {\ensuremath{\mathrm{d}\hspace{-0.1em}p/p}\xspace}

\newcommand{\dedx}{\ensuremath{\mathrm{d}\hspace{-0.1em}E/\mathrm{d}x}\xspace}


\def\dllkpi     {\ensuremath{\mathrm{DLL}_{\kaon\pion}}\xspace}
\def\dllppi     {\ensuremath{\mathrm{DLL}_{\proton\pion}}\xspace}
\def\dllepi     {\ensuremath{\mathrm{DLL}_{\electron\pion}}\xspace}
\def\dllmupi    {\ensuremath{\mathrm{DLL}_{\mmu\pi}}\xspace}

\def\mphi       {\mbox{$\phi$}\xspace}
\def\mtheta     {\mbox{$\theta$}\xspace}
\def\ctheta     {\mbox{$\cos\theta$}\xspace}
\def\stheta     {\mbox{$\sin\theta$}\xspace}
\def\ttheta     {\mbox{$\tan\theta$}\xspace}

\def\degrees{\ensuremath{^{\circ}}\xspace}
\def\krad {\ensuremath{\rm \,krad}\xspace}
\def\mrad{\ensuremath{\rm \,mrad}\xspace}
\def\rad{\ensuremath{\rm \,rad}\xspace}

\def\betastar {\ensuremath{\beta^*}}
\newcommand{\lum} {\ensuremath{\mathcal{L}}\xspace}
\newcommand{\intlum}[1]{\ensuremath{\int\lum=#1\xspace}}  


\def\evtgen     {\mbox{\textsc{EvtGen}}\xspace}
\def\pythia     {\mbox{\textsc{Pythia}}\xspace}
\def\fluka      {\mbox{\textsc{Fluka}}\xspace}
\def\tosca      {\mbox{\textsc{Tosca}}\xspace}
\def\ansys      {\mbox{\textsc{Ansys}}\xspace}
\def\spice      {\mbox{\textsc{Spice}}\xspace}
\def\garfield   {\mbox{\textsc{Garfield}}\xspace}
\def\geant      {\mbox{\textsc{Geant3}}\xspace}
\def\hepmc      {\mbox{\textsc{HepMC}}\xspace}
\def\gauss      {\mbox{\textsc{Gauss}}\xspace}
\def\gaudi      {\mbox{\textsc{Gaudi}}\xspace}
\def\boole      {\mbox{\textsc{Boole}}\xspace}
\def\brunel     {\mbox{\textsc{Brunel}}\xspace}
\def\davinci    {\mbox{\textsc{DaVinci}}\xspace}
\def\erasmus    {\mbox{\textsc{Erasmus}}\xspace}
\def\moore      {\mbox{\textsc{Moore}}\xspace}
\def\ganga      {\mbox{\textsc{Ganga}}\xspace}
\def\dirac      {\mbox{\textsc{Dirac}}\xspace}
\def\root       {\mbox{\textsc{Root}}\xspace}
\def\roofit     {\mbox{\textsc{RooFit}}\xspace}
\def\pyroot     {\mbox{\textsc{PyRoot}}\xspace}
\def\photos     {\mbox{\textsc{Photos}}\xspace}

\def\cpp        {\mbox{\textsc{C\raisebox{0.1em}{{\footnotesize{++}}}}}\xspace}
\def\python     {\mbox{\textsc{Python}}\xspace}
\def\ruby       {\mbox{\textsc{Ruby}}\xspace}
\def\fortran    {\mbox{\textsc{Fortran}}\xspace}
\def\svn        {\mbox{\textsc{SVN}}\xspace}

\def\kbytes     {\ensuremath{{\rm \,kbytes}}\xspace}
\def\kbsps      {\ensuremath{{\rm \,kbytes/s}}\xspace}
\def\kbits      {\ensuremath{{\rm \,kbits}}\xspace}
\def\kbsps      {\ensuremath{{\rm \,kbits/s}}\xspace}
\def\mbsps      {\ensuremath{{\rm \,Mbits/s}}\xspace}
\def\mbytes     {\ensuremath{{\rm \,Mbytes}}\xspace}
\def\mbps       {\ensuremath{{\rm \,Mbyte/s}}\xspace}
\def\mbsps      {\ensuremath{{\rm \,Mbytes/s}}\xspace}
\def\gbsps      {\ensuremath{{\rm \,Gbits/s}}\xspace}
\def\gbytes     {\ensuremath{{\rm \,Gbytes}}\xspace}
\def\gbsps      {\ensuremath{{\rm \,Gbytes/s}}\xspace}
\def\tbytes     {\ensuremath{{\rm \,Tbytes}}\xspace}
\def\tbpy       {\ensuremath{{\rm \,Tbytes/yr}}\xspace}

\def\dst        {DST\xspace}


\def\nonn {\ensuremath{\rm {\it{n^+}}\mbox{-}on\mbox{-}{\it{n}}}\xspace}
\def\ponn {\ensuremath{\rm {\it{p^+}}\mbox{-}on\mbox{-}{\it{n}}}\xspace}
\def\nonp {\ensuremath{\rm {\it{n^+}}\mbox{-}on\mbox{-}{\it{p}}}\xspace}
\def\cvd  {CVD\xspace}
\def\mwpc {MWPC\xspace}
\def\gem  {GEM\xspace}

\def\tell1  {TELL1\xspace}
\def\ukl1   {UKL1\xspace}
\def\beetle {Beetle\xspace}
\def\otis   {OTIS\xspace}
\def\croc   {CROC\xspace}
\def\carioca {CARIOCA\xspace}
\def\dialog {DIALOG\xspace}
\def\sync   {SYNC\xspace}
\def\cardiac {CARDIAC\xspace}
\def\gol    {GOL\xspace}
\def\vcsel  {VCSEL\xspace}
\def\ttc    {TTC\xspace}
\def\ttcrx  {TTCrx\xspace}
\def\hpd    {HPD\xspace}
\def\pmt    {PMT\xspace}
\def\specs  {SPECS\xspace}
\def\elmb   {ELMB\xspace}
\def\fpga   {FPGA\xspace}
\def\plc    {PLC\xspace}
\def\rasnik {RASNIK\xspace}
\def\elmb   {ELMB\xspace}
\def\can    {CAN\xspace}
\def\lvds   {LVDS\xspace}
\def\ntc    {NTC\xspace}
\def\adc    {ADC\xspace}
\def\led    {LED\xspace}
\def\ccd    {CCD\xspace}
\def\hv     {HV\xspace}
\def\lv     {LV\xspace}
\def\pvss   {PVSS\xspace}
\def\cmos   {CMOS\xspace}
\def\fifo   {FIFO\xspace}
\def\ccpc   {CCPC\xspace}

\def\cfourften     {\ensuremath{\rm C_4 F_{10}}\xspace}
\def\cffour        {\ensuremath{\rm CF_4}\xspace}
\def\cotwo         {\ensuremath{\rm CO_2}\xspace} 
\def\csixffouteen  {\ensuremath{\rm C_6 F_{14}}\xspace} 
\def\mgftwo     {\ensuremath{\rm Mg F_2}\xspace} 
\def\siotwo     {\ensuremath{\rm SiO_2}\xspace} 

\newcommand{\eg}{\mbox{\itshape e.g.}\xspace}
\newcommand{\ie}{\mbox{\itshape i.e.}}
\newcommand{\etal}{{\slshape et al.\/}\xspace}
\newcommand{\etc}{\mbox{\itshape etc.}\xspace}
\newcommand{\cf}{\mbox{\itshape cf.}\xspace}
\newcommand{\ffp}{\mbox{\itshape ff.}\xspace}

\def\Kbar    {\ensuremath{\overline{K}{}^0}\xspace}
\def\KS      {\ensuremath{K^0_S}\xspace} 
\def\KSb     {\ensuremath{\Kbar^0_S}\xspace} 
\def\KL      {\ensuremath{K^0_L}\xspace} 








\title{ 
Evidence for the decay \mbox{\boldmath$\Bz\to p\bar{p}\piz$}}



\noaffiliation
\affiliation{University of the Basque Country UPV/EHU, 48080 Bilbao}
\affiliation{Beihang University, Beijing 100191}
\affiliation{Brookhaven National Laboratory, Upton, New York 11973}
\affiliation{Budker Institute of Nuclear Physics SB RAS, Novosibirsk 630090}
\affiliation{Faculty of Mathematics and Physics, Charles University, 121 16 Prague}
\affiliation{Chonnam National University, Kwangju 660-701}
\affiliation{University of Cincinnati, Cincinnati, Ohio 45221}
\affiliation{Deutsches Elektronen--Synchrotron, 22607 Hamburg}
\affiliation{University of Florida, Gainesville, Florida 32611}
\affiliation{Key Laboratory of Nuclear Physics and Ion-beam Application (MOE) and Institute of Modern Physics, Fudan University, Shanghai 200443}
\affiliation{II. Physikalisches Institut, Georg-August-Universit\"at G\"ottingen, 37073 G\"ottingen}
\affiliation{SOKENDAI (The Graduate University for Advanced Studies), Hayama 240-0193}
\affiliation{Gyeongsang National University, Chinju 660-701}
\affiliation{Hanyang University, Seoul 133-791}
\affiliation{University of Hawaii, Honolulu, Hawaii 96822}
\affiliation{High Energy Accelerator Research Organization (KEK), Tsukuba 305-0801}
\affiliation{J-PARC Branch, KEK Theory Center, High Energy Accelerator Research Organization (KEK), Tsukuba 305-0801}
\affiliation{Forschungszentrum J\"{u}lich, 52425 J\"{u}lich}
\affiliation{IKERBASQUE, Basque Foundation for Science, 48013 Bilbao}
\affiliation{Indian Institute of Science Education and Research Mohali, SAS Nagar, 140306}
\affiliation{Indian Institute of Technology Guwahati, Assam 781039}
\affiliation{Indian Institute of Technology Hyderabad, Telangana 502285}
\affiliation{Indian Institute of Technology Madras, Chennai 600036}
\affiliation{Indiana University, Bloomington, Indiana 47408}
\affiliation{Institute of High Energy Physics, Chinese Academy of Sciences, Beijing 100049}
\affiliation{Institute of High Energy Physics, Vienna 1050}
\affiliation{Institute for High Energy Physics, Protvino 142281}
\affiliation{INFN - Sezione di Napoli, 80126 Napoli}
\affiliation{INFN - Sezione di Torino, 10125 Torino}
\affiliation{Advanced Science Research Center, Japan Atomic Energy Agency, Naka 319-1195}
\affiliation{J. Stefan Institute, 1000 Ljubljana}
\affiliation{Institut f\"ur Experimentelle Teilchenphysik, Karlsruher Institut f\"ur Technologie, 76131 Karlsruhe}
\affiliation{Kennesaw State University, Kennesaw, Georgia 30144}
\affiliation{King Abdulaziz City for Science and Technology, Riyadh 11442}
\affiliation{Korea Institute of Science and Technology Information, Daejeon 305-806}
\affiliation{Korea University, Seoul 136-713}
\affiliation{Kyungpook National University, Daegu 702-701}
\affiliation{LAL, Univ. Paris-Sud, CNRS/IN2P3, Universit\'{e} Paris-Saclay, Orsay}
\affiliation{\'Ecole Polytechnique F\'ed\'erale de Lausanne (EPFL), Lausanne 1015}
\affiliation{P.N. Lebedev Physical Institute of the Russian Academy of Sciences, Moscow 119991}
\affiliation{Faculty of Mathematics and Physics, University of Ljubljana, 1000 Ljubljana}
\affiliation{Ludwig Maximilians University, 80539 Munich}
\affiliation{Luther College, Decorah, Iowa 52101}
\affiliation{University of Malaya, 50603 Kuala Lumpur}
\affiliation{University of Maribor, 2000 Maribor}
\affiliation{Max-Planck-Institut f\"ur Physik, 80805 M\"unchen}
\affiliation{School of Physics, University of Melbourne, Victoria 3010}
\affiliation{University of Mississippi, University, Mississippi 38677}
\affiliation{University of Miyazaki, Miyazaki 889-2192}
\affiliation{Moscow Physical Engineering Institute, Moscow 115409}
\affiliation{Moscow Institute of Physics and Technology, Moscow Region 141700}
\affiliation{Graduate School of Science, Nagoya University, Nagoya 464-8602}
\affiliation{Kobayashi-Maskawa Institute, Nagoya University, Nagoya 464-8602}
\affiliation{Universit\`{a} di Napoli Federico II, 80055 Napoli}
\affiliation{Nara Women's University, Nara 630-8506}
\affiliation{National Central University, Chung-li 32054}
\affiliation{National United University, Miao Li 36003}
\affiliation{Department of Physics, National Taiwan University, Taipei 10617}
\affiliation{H. Niewodniczanski Institute of Nuclear Physics, Krakow 31-342}
\affiliation{Nippon Dental University, Niigata 951-8580}
\affiliation{Niigata University, Niigata 950-2181}
\affiliation{Novosibirsk State University, Novosibirsk 630090}
\affiliation{Pacific Northwest National Laboratory, Richland, Washington 99352}
\affiliation{Panjab University, Chandigarh 160014}
\affiliation{Peking University, Beijing 100871}
\affiliation{University of Pittsburgh, Pittsburgh, Pennsylvania 15260}
\affiliation{Theoretical Research Division, Nishina Center, RIKEN, Saitama 351-0198}
\affiliation{University of Science and Technology of China, Hefei 230026}
\affiliation{Seoul National University, Seoul 151-742}
\affiliation{Showa Pharmaceutical University, Tokyo 194-8543}
\affiliation{Soongsil University, Seoul 156-743}
\affiliation{University of South Carolina, Columbia, South Carolina 29208}
\affiliation{Stefan Meyer Institute for Subatomic Physics, Vienna 1090}
\affiliation{Sungkyunkwan University, Suwon 440-746}
\affiliation{School of Physics, University of Sydney, New South Wales 2006}
\affiliation{Department of Physics, Faculty of Science, University of Tabuk, Tabuk 71451}
\affiliation{Tata Institute of Fundamental Research, Mumbai 400005}
\affiliation{Department of Physics, Technische Universit\"at M\"unchen, 85748 Garching}
\affiliation{Toho University, Funabashi 274-8510}
\affiliation{Department of Physics, Tohoku University, Sendai 980-8578}
\affiliation{Earthquake Research Institute, University of Tokyo, Tokyo 113-0032}
\affiliation{Department of Physics, University of Tokyo, Tokyo 113-0033}
\affiliation{Tokyo Institute of Technology, Tokyo 152-8550}
\affiliation{Tokyo Metropolitan University, Tokyo 192-0397}
\affiliation{Virginia Polytechnic Institute and State University, Blacksburg, Virginia 24061}
\affiliation{Wayne State University, Detroit, Michigan 48202}
\affiliation{Yamagata University, Yamagata 990-8560}
\affiliation{Yonsei University, Seoul 120-749}
 \author{B.~Pal}\affiliation{Brookhaven National Laboratory, Upton, New York 11973} 
  \author{I.~Adachi}\affiliation{High Energy Accelerator Research Organization (KEK), Tsukuba 305-0801}\affiliation{SOKENDAI (The Graduate University for Advanced Studies), Hayama 240-0193} 
  \author{K.~Adamczyk}\affiliation{H. Niewodniczanski Institute of Nuclear Physics, Krakow 31-342} 
  \author{H.~Aihara}\affiliation{Department of Physics, University of Tokyo, Tokyo 113-0033} 
  \author{D.~M.~Asner}\affiliation{Brookhaven National Laboratory, Upton, New York 11973} 
  \author{H.~Atmacan}\affiliation{University of South Carolina, Columbia, South Carolina 29208} 
  \author{V.~Aulchenko}\affiliation{Budker Institute of Nuclear Physics SB RAS, Novosibirsk 630090}\affiliation{Novosibirsk State University, Novosibirsk 630090} 
  \author{T.~Aushev}\affiliation{Moscow Institute of Physics and Technology, Moscow Region 141700} 
  \author{R.~Ayad}\affiliation{Department of Physics, Faculty of Science, University of Tabuk, Tabuk 71451} 
  \author{V.~Babu}\affiliation{Tata Institute of Fundamental Research, Mumbai 400005} 
% ...
\collaboration{The Belle Collaboration}


\begin{abstract}
\noindent
We report a search for the charmless baryonic decay $\Bz\to p\bar{p}\piz$  with a data sample corresponding
to an integrated luminosity of 711~$\rm fb^{-1}$ containing $(772\pm 10)\times 10^6$ $\B\Bb$ pairs. The data was collected by
the Belle experiment running on the $\Y4S$ resonance at the KEKB $\epem$ collider.  We measure a branching fraction $\mathcal{B}(\Bd\to p\bar{p}\pi^0)= (5.0\pm1.8\pm0.6 )\times 10^{-7}$, 
where the first uncertainty is statistical and the second is systematic. 
The signal has a significance of 3.1 standard deviations and constitutes the first evidence for this decay mode. We also search for the intermediate two-body decays $\Bd\to\Delta^+\bar{p}$ and  $\Bd\to\bar{\Delta}^-p$, and set an upper limit  on the branching fraction: $\mathcal{B}(\Bd\to \Delta^+\bar{p})+\mathcal{B}(\Bd\to\bar{\Delta}^-p)<1.6\times10^{-6}$ at 90\% confidence level.

\end{abstract}


\maketitle
\tighten
The first observed charmless baryonic $\B$ decay was $\Bp\to p\bar{p}\Kp$~\cite{Abe:2002ds}. Following this first observation, many other charmless baryonic $B$ decays have been found~\cite{PDG}. Except for $\Bp\to p\L \piz ~{\rm and} ~p\L\gamma$ decays, all the channels reported to date are entirely reconstructed from charged particles  in the final state.    A noticeable hierarchy is also observed in the branching fractions of these decays: three-body decays are usually more frequent than their two-body counterparts but less frequent than four-body decays~\cite{Bevan:2014iga, Cheng:2014qxa}. This phenomenon can be understood in terms of the so-called ``threshold effect," which refers to the fact that the $\B$ meson prefers to decay into a di-baryon pair with low invariant mass accompanied by a fast recoil meson~\cite{Bevan:2014iga, Hou:2000bz, Chen:2008sw}. 
This peaking behavior was  unexpected, and has led to various speculations about possible mechanisms~\cite{Bevan:2014iga}. 
Studying additional three-body baryonic decays
might provide a better understanding of the dynamics of $\B$ decays and the aforementioned threshold-effect. These decays are also useful for  \CP violation studies. 

This paper reports a search for a three-body charmless baryonic $\Bd$ decays  to the $p\bar{p}\piz$ final state~\cite{charge-conjugate} using a data set corresponding to an integrated luminosity of 711~$\rm fb^{-1}$ collected with the Belle detector~\cite{Belle}  at the \FourS resonance at the KEKB asymmetric-energy $\epem$  (3.5 on 8.0 GeV) collider~\cite{KEKB}.
So far, the decay $\Bz\to p\bar{p}\piz$ has not been studied by any experiment. No theoretical prediction for the branching fraction of
this process is yet available. 
A glance at the known branching fractions for  $\B$ decays~\cite{PDG} shows  the three-body charmless baryonic decays to occur in the several times $10^{-6}$ range, indicating that the discovery of the mode $\Bz\to p\bar{p}\piz$ might be possible 
with the currently available data set.  


The Belle detector is a large-solid-angle magnetic spectrometer consisting
of a  silicon vertex detector (SVD), a 50-layer central
drift chamber (CDC), an array of aerogel threshold Cherenkov counters (ACC),
a barrel-like arrangement of time-of-flight scintillation counters (TOF),
and an electromagnetic calorimeter (ECL) comprising  CsI(Tl) crystals. 
These detector components are 
located inside a superconducting solenoid coil that provides a 1.5~T
magnetic field.  An iron flux-return located outside the coil 
is instrumented (KLM) to detect $K_L^0$ mesons and to identify muons.
Two inner detector configurations were used: a 2.0 cm radius beampipe
and a three-layer SVD were used for the first 
$152\times 10^6$ $\BBbar$ pairs of data, 
while a 1.5 cm radius beampipe, a four-layer SVD, and a 
small-cell inner drift chamber were used for the remaining
$620\times10^6$ $\BBbar$ pairs of data.
The detector is described in detail in Ref.~\cite{Belle}.
Event selection requirements are optimized using Monte Carlo (MC) simulations.
MC events are generated using \evtgen~\cite{Lange:2001uf}, and the detector 
response is modeled using \geant~\cite{geant3}.  Final-state radiation is 
taken into account using the \photos package~\cite{Golonka:2005pn}.  

The reconstruction of $\Bz\to p\bar{p}\piz$ proceeds by first reconstructing 
$\piz\to\gamma\gamma$ 
candidates. An ECL cluster not matched to any track in the CDC is identified as a photon candidate. 
Such candidates are required to have an energy greater than 50~MeV in the barrel region 
and greater than 100~MeV in the end-cap regions, where the barrel region covers the polar 
angle $32^{\circ} < \theta < 130^{\circ}$ and the end-cap regions cover the ranges 
$12^{\circ} < \theta < 32^{\circ}$ and $130^{\circ}<\theta <157^{\circ}$. 
To reject showers produced by neutral hadrons, the  energy deposited in the $3\times3$ array of ECL crystals centered on the crystal with the highest energy must exceed 80\% of the energy deposited in the corresponding $5\times5$ array of crystals.
We require 
that the $\gamma\gamma$ invariant mass be within  $20 \mevcc$ (about $3.5\sigma$ 
in resolution) of the $\piz$ mass~\cite{PDG}. 
To improve the $\pi^0$ momentum resolution, 
we perform a mass-constrained fit and require that the resulting $\chi^2$ be less than~30. 
This requirement is relatively loose, retaining  more than 99\% of candidates.

We subsequently combine $\piz$ candidates with two oppositely charged  tracks, identified as  a proton-antiproton pair.
Such tracks  are identified using  requirements on the distance of closest approach with respect to the interaction point  along the $z$ axis (antiparallel to the $e^+$ beam) of $|dz|< 3.0$~cm, and in the transverse plane of $dr<0.3$~cm. In addition, charged tracks are required to have a minimum number of SVD hits 
($>2$ along  the  $z$ axis and $>1$ in the transverse direction). 
Particle identification is achieved using information from the CDC,  the TOF, and the ACC subdetectors. This information is combined to form a hadron likelihood $\mathcal{L}_h$; a charged track with likelihood ratios of $\mathcal{L}_p/(\mathcal{L}_{p}+\mathcal{L}_K)> 0.9$ and $\mathcal{L}_p/(\mathcal{L}_{p}+\mathcal{L}_{\pi}) > 0.9$ is regarded as a proton or antiproton.
Furthermore, we reject 
tracks consistent with either the electron or muon hypothesis. The proton identification efficiency is 75\% and the probability for a kaon (pion) to be misidentified as a proton is 6\%~(2\%).

Candidate $\Bz$ mesons are identified using the beam-energy-constrained mass,
$M_{\rm bc}=\sqrt{E_{\rm beam}^2-|\vec{p}_{B} c|^2}/c^2$,
and the energy difference
$\Delta E = E_B-E_{\rm beam}$,
where $E_{\rm beam}$ is the beam energy, and $E_B$ and $\vec{p}_B$ are the reconstructed  energy and  momentum, respectively, of the $\Bz$ candidate. All quantities are evaluated in the  center-of-mass (CM) frame.  To improve the $M_{\rm bc}$ resolution,  the momentum $\vec{p}_B$ is calculated as
$\vec{p}_B= \vec{p}_{p}+\vec{p}_{\bar{p}}+\frac{\vec{p}_{\piz}}{|\vec{p}_{\piz}|}\sqrt{(E_{\rm beam}-E_{p}-E_{\bar{p}})^2/c^2-m^2_{\piz}c^2}$,
where $m_{\piz}$ is the nominal $\piz$ mass~\cite{PDG}, $E_{h}$  and $\vec{p}_{h}$ are the energy and momentum of the  hadron $h$~($h=p, ~\bar{p}, ~\piz$). In addition, a vertex fit is performed to the charged tracks to form a  $\Bz$ vertex. We require that the $\chi^2$ from the fit be less than 200.  Events with $M_{\rm bc}>5.25~\gevcc$ and $-0.20~\gev<\Delta E < 0.15~\gev$ are retained for further analysis. The signal yield is calculated in a smaller region $M_{\rm bc}\in (5.272,5.286)$~\gevcc and $\Delta E\in (-0.12,+0.06)$~\gev.  
In order to reject contributions from charmonium states ($e.g.$, $\etac$, $\jpsi$, $\psi(2S)$, $\chi_{c0}$,  $\chi_{c1}$ and $\chi_{c2}$), we apply a ``charmonium veto" and exclude the regions of $2.850~ \gevcc < m(p\bar{p}) < 3.128~\gevcc$ and $3.315~\gevcc < m(p\bar{p}) < 3.735~\gevcc$ from the event sample. 

Charmless hadronic decays suffer from large amount of continuum background,   arising from light quark production ($\epem\to\qqbar,~q=u,d,s,c$).  To suppress this background, we use a multivariate analyzer based on a neural network (NN)~\cite{Feindt:2006pm} that distinguishes jet-like continuum events from more spherical $\BBbar$ events. The NN uses the following input variables: the cosine of the angle between the thrust axis~\cite{Brandt:1964sa}  of the $\Bz$ candidate and the thrust axis of the rest of the event; the cosine of the angle between the $\Bz$ thrust axis and the $+z$ axis; the cosine of the angle between the $+z$ axis and the $\Bz$  candidate flight direction; a set of 18 modified Fox-Wolfram moments~\cite{SFW}; the ratio of the second to zeroth (unmodified) Fox-Wolfram moments; the separation along the $z$ axis between the two $\B$ vertices; and the $\B$-flavor tagging information~\cite{Kakuno:2004cf}. 
All but for the last two quantities are evaluated in the  CM frame.
The NN is trained using MC simulated signal events 
and $\qqbar$ background events.
The NN generates a single output variable ($C_{\rm NN}$) that ranges from $-1$ for background-like events to $+1$ for signal-like events. 
We require $C_{\rm NN}>-0.5$, which rejects 
approximately 86\% of the $\qqbar$ background while retaining 94\%
of the signal. We then translate $C_{\rm NN}$ to a new
variable 
\begin{linenomath}
\begin{equation}
C'_{\rm NN} = \ln\left(\frac{C_{\rm NN}-C^{\rm min}_{\rm NN}}
{C^{\rm max}_{\rm NN}-C_{\rm NN}}\right),
\end{equation} 
\end{linenomath}
where $C^{\rm min}_{\rm NN}= -0.5$ and $C_{\rm NN}^{\rm max}=1.0$.
 This translation is advantageous as the $C'_{\rm NN}$ distribution
for both signal and background is well described by a  sum of
Gaussian functions. 

After applying all selection criteria, approximately 7\% of the events have multiple $\Bd$ candidates. 
For these events, we retain the candidate having the smallest sum of $\chi^2$ values obtained from 
the $\piz\to\gamma\gamma$ mass-constrained fit and the $\Bd$ vertex-constrained fit. 
According to MC simulation, this criterion selects the correct $\Bd$ candidate in 83\% of multiple-candidate events.

We measure the signal yield by performing an unbinned extended
maximum likelihood fit to the variables $M_{\rm bc}$, $\Delta E$,
and $C^{\prime}_{\rm NN}$. The likelihood function is defined as
\begin{linenomath}
\begin{equation}
\mathcal{L} = e^{-\sum_j Y_j}
\prod_i^N \left( \sum_j Y_j \mathcal{P}_j(M_{\rm bc}^i, \Delta E^i, C'^i_{\rm NN} )\right),
\end{equation} 
\end{linenomath}
where $Y_j$ is the yield of  component~$j$;
$\mathcal{P}_j(M_{\rm bc}^i, \Delta E^i, C'^i_{\rm NN})$ is the
probability density function (PDF) 
 of component $j$ for event $i$;
$j$ runs over all signal and  background components;
and $i$ runs over all events in the sample~($N$).
The background components consist of continuum events,  $b\to c$ (generic $\B$) processes, and rare charmless  processes. 
The latter two backgrounds are small compared to the continuum events and are studied using MC simulations. 
The rare charmless background shows a peaking structure in the $M_{\rm bc}$ distribution, most of which arises from 
$\Bp\to p\bar{p}\rho^+$ decays. As correlations among the variables $M_{\rm bc}$, $\Delta E$,
and $C^{\prime}_{\rm NN}$ are found to be small, the three-dimensional PDFs
$\mathcal{P}_j(M_{\rm bc}^i, \Delta E^i, C'^i_{\rm NN})$ are factorized 
into the product of separate one-dimensional PDFs.

The PDF of signal events consists of two parts: one for candidates that are 
correctly reconstructed, and one for those incorrectly reconstructed, 
$i.e.$, at least one  daughter originates from the other (tag-side) $B$. 
For the former case, the $M_{\rm bc}$ and $\Delta E$ distributions are modeled with  Gaussian and Crystal Ball (CB)~\cite{Skwarnicki:1986xj} functions, respectively, while  the $C'_{\rm NN}$ distribution is modeled with a sum of Gaussian and bifurcated Gaussian functions having a common mean.  
The peak positions and resolutions of the $M_{\rm bc}$, $\Delta E$, and $C'_{\rm NN}$ PDFs are adjusted to account for data-MC differences observed in a high-statistics control sample of $\Bd\to\Dzb(\to \Kp\pim)\piz$ decays. 
For the latter case, the 
correlated two-dimensional $M_{\rm bc}$-$\Delta E$ distribution 
is modeled with a non-parametric PDF~\cite{Cranmer:2000du}, and the $C'_{\rm NN}$ component is modeled with a Gaussian function.
The fraction of incorrectly reconstructed  decays ($\sim 4\%$ in the signal region) is taken from MC simulation. For the rare charmless  background, the  $C'_{\rm NN}$ component is modeled with a bifurcated Gaussian function. The  $M_{\rm bc}$ and $\Delta E$  components are modeled by a joint two-dimensional non-parametric PDF.  We model the $M_{\rm bc}$, $\Delta E$, and  $C'_{\rm NN}$ distributions of continuum background
with an ARGUS~\cite{Albrecht:1990am} function having its endpoint fixed to $5.29~\gevcc$,  a 
first-order polynomial,  and a sum of two Gaussians having a common mean,  respectively.
For the generic $\B$ background, we use a bifurcated Gaussian function to model the $C'_{\rm NN}$ shape, while the 
similar shapes as of continuum
 background are used to model the $M_{\rm bc}$ and $\Delta E$ distributions. 
In addition to the fitted yields $Y_j$, all shape parameters for continuum background are also floated. 
All other parameters are fixed to the corresponding MC values. 

\begin{figure}[h!t!p!]
\begin{center}
    \includegraphics[width=0.5\textwidth]{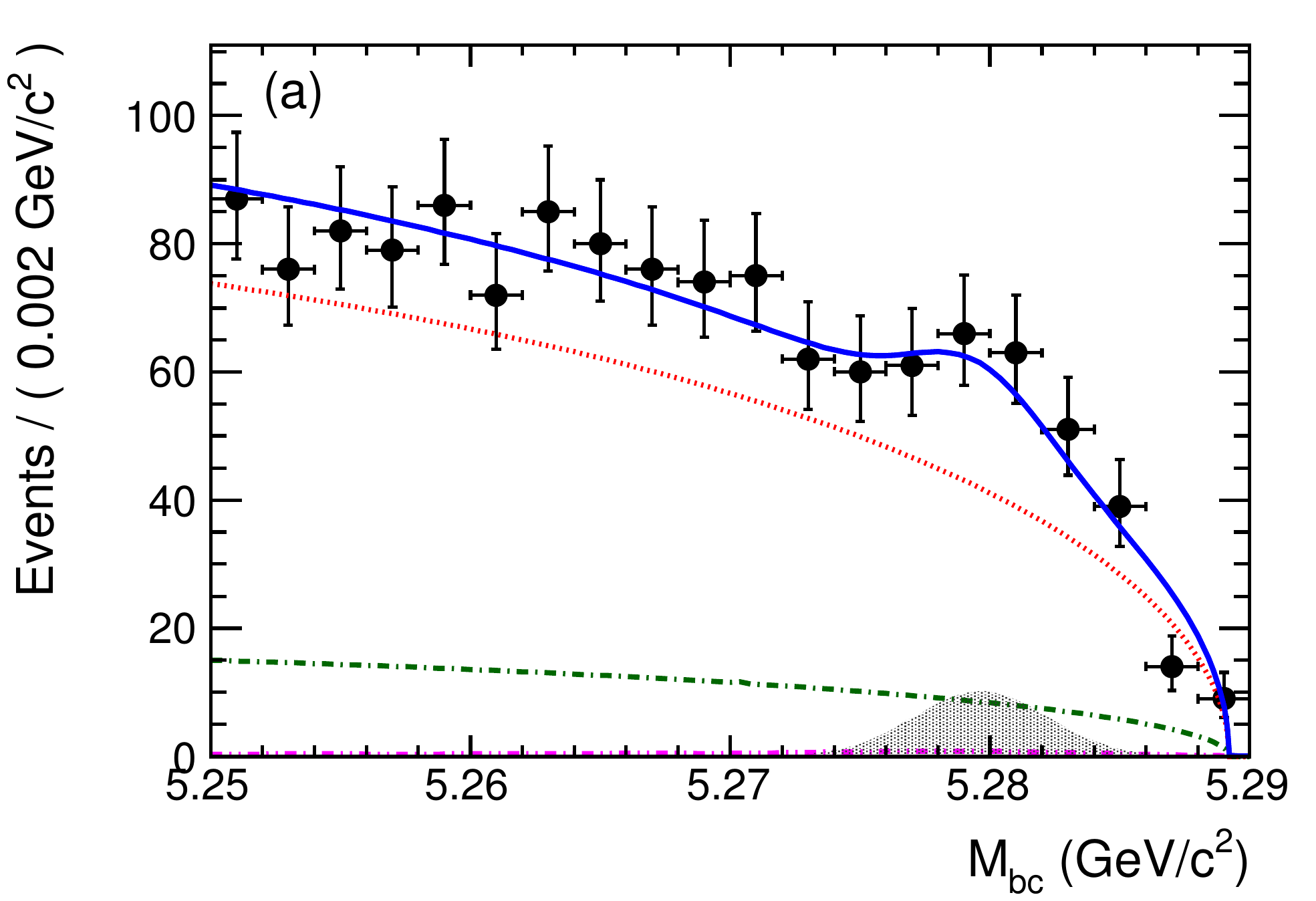}
    \includegraphics[width=0.5\textwidth]{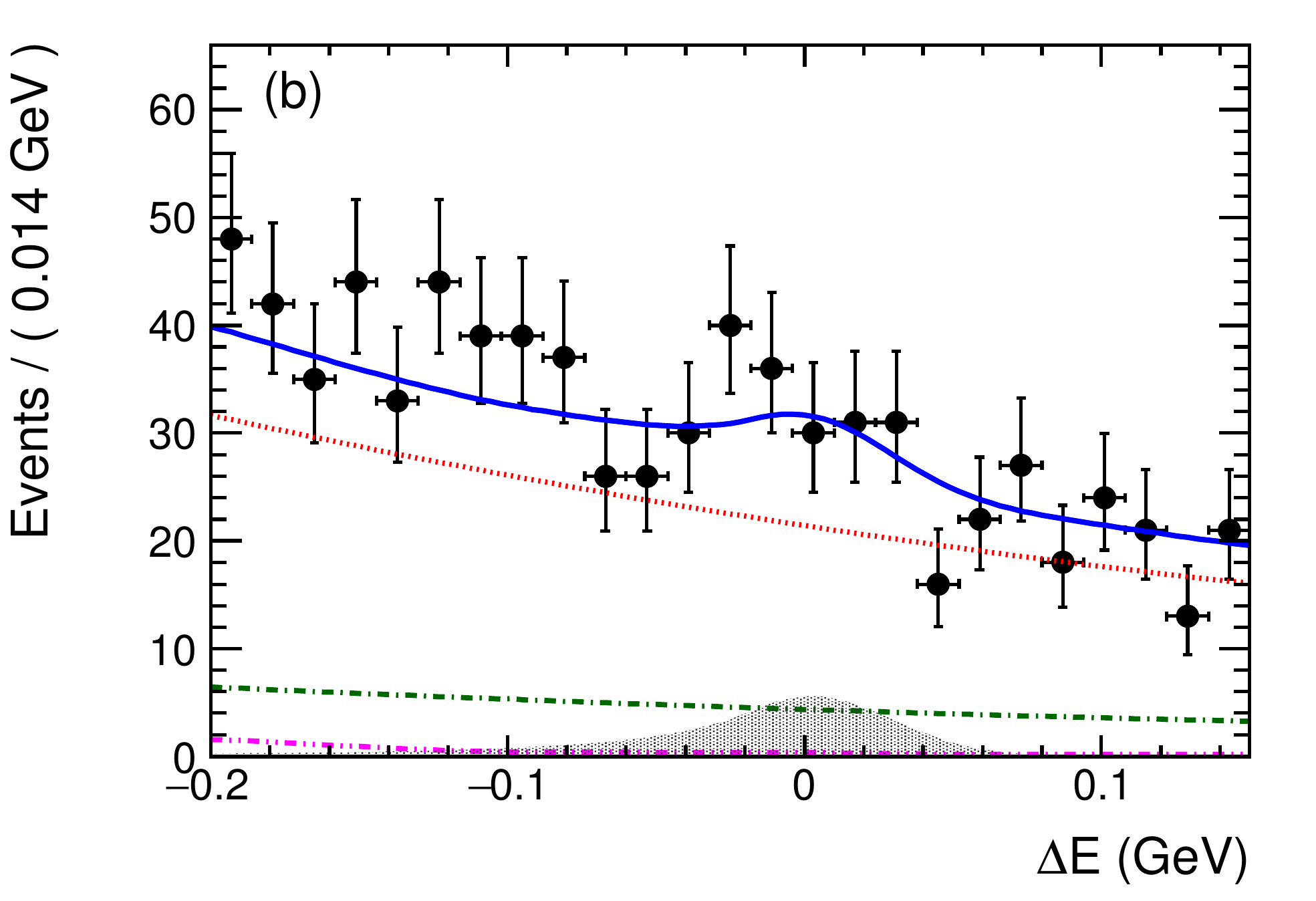}
    \includegraphics[width=0.5\textwidth]{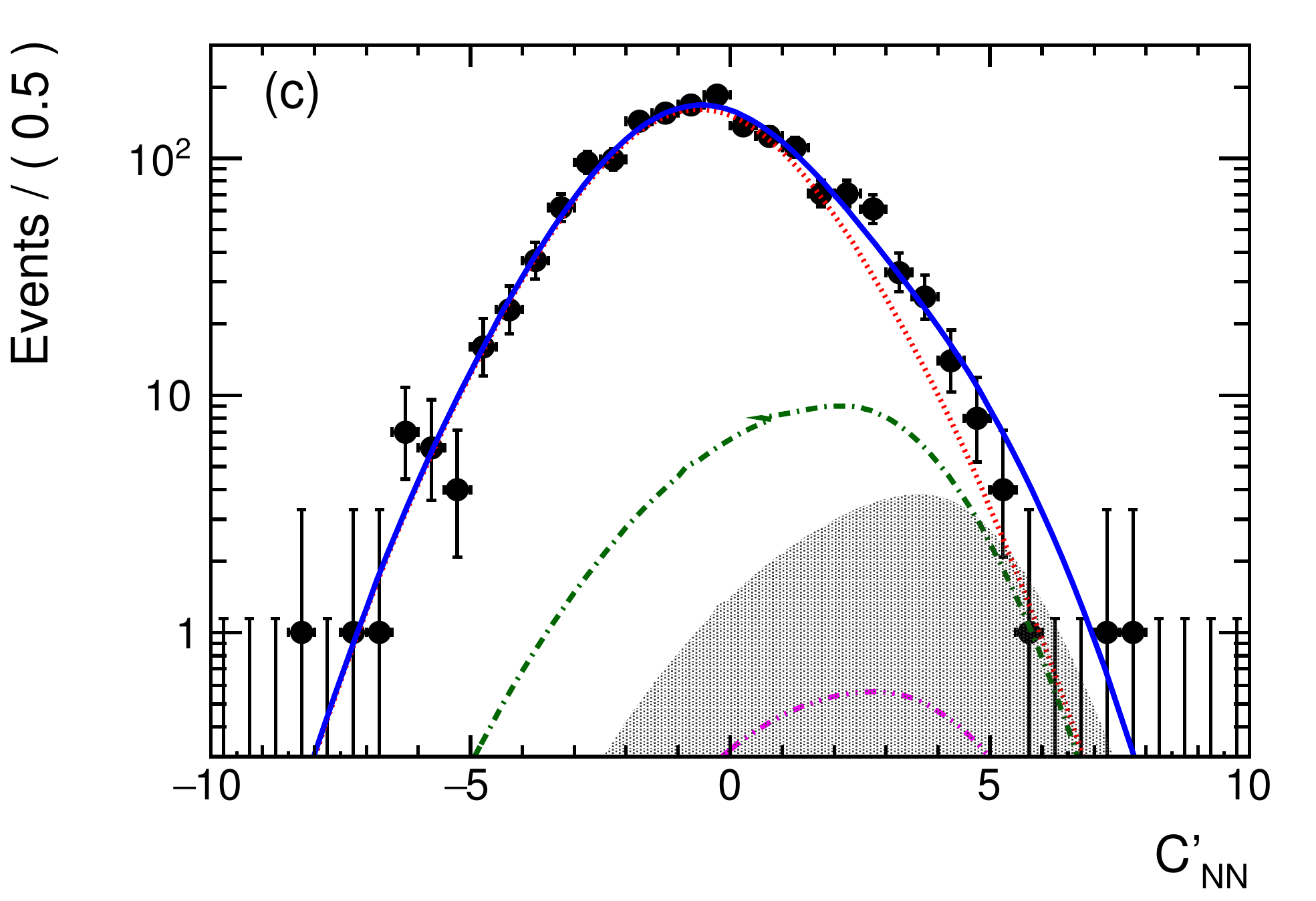}
\end{center}\label{fig:dat_signal}
\vskip -0.5cm
\caption{\small Projection of the 3D fit to the data: (a) $M_{\rm bc}$ in $-0.12 < \Delta E <0.06$ GeV and $C'_{\rm NN}>1.0$,  (b) $\Delta E$ in $5.272< M_{\rm bc} < 5.286$ GeV and $C'_{\rm NN}>1.0$, and (c) $C'_{\rm NN}$ in $5.272< M_{\rm bc} < 5.286$ GeV and $-0.12< \Delta E <0.06$ GeV. Points with  error bars are data, shaded areas represent the signal, (red) dotted curves denote the continuum background, (green) dot-dashed curves the generic $\B$ background, (magenta) dot-dot-dashed curves the rare charmless  background, and  (blue) curves show the total contribution.  The $\chi^2$/(number of bins) values of these fit projections are 0.60, 0.83, and  0.91, respectively, which indicates that the fit gives a good description of the data.}
\end{figure}

The projections of the fit are shown in Fig.~\ref{fig:dat_signal}. 
From the fit, we extract $40.5\pm14.2$ signal events, $1490.3\pm34.5$ continuum,  $100.6\pm35.0$ generic $\B$, and $6.5\pm10.1$ rare charmless background events in the $M_{\rm bc} - \Delta E$ signal region. The resulting branching fraction is calculated as
\begin{linenomath}
\begin{equation}
\mathcal{B}(\Bd\to p\bar{p}\pi^0)=\frac{Y_{\rm sig}}{N_{\BBbar}\times \varepsilon }, \label{eq:br}
\end{equation}
\end{linenomath}
where $Y_{\rm sig}$ represents the extracted signal yield, $N_{\B\Bb} = (772\pm11)\times10^6$ is the total number of $\BBbar$ events, $\varepsilon = (10.53\pm0.04)\%$ is the reconstruction efficiency. The efficiency is corrected to account for possible differences in particle identification (PID) 
and $\piz$ detection efficiencies between data and simulations.  In
Eq.~(\ref{eq:br}) we assume equal production of $\Bd\Bdb$ and $\Bp\Bm$
pairs at the $\FourS$ resonance. The result is
\begin{linenomath}
\begin{equation*}
\mathcal{B}(\Bd\to p\bar{p}\pi^0)= (5.0\pm1.8\pm0.6 )\times10^{-7},
\end{equation*}
\end{linenomath}
where the first uncertainty is statistical and the second
is systematic. This is the first measurement of this branching fraction.

The signal significance is calculated as  $\sqrt{-2\ln(\mathcal{L}_0/\mathcal{L}_{\rm max})}$, where $\mathcal{L}_0$ is the likelihood value when the signal yield is fixed to zero, and $\mathcal{L}_{\rm max}$ is the likelihood value of the nominal fit. To include systematic uncertainties in the significance, we convolve the likelihood distribution with a Gaussian function whose width is set to the total systematic uncertainty that affects the signal yield. The resulting significance is 3.1 standard deviations. Thus, our measurement constitutes the first evidence for this decay mode. 

The systematic uncertainty in $\BF(\Bz\to p\bar{p}\piz)$ arises from several sources, as listed in Table~\ref{tab:sys1}. The uncertainty due to the fixed parameters in the PDF is estimated by varying them individually according to their statistical uncertainties. 
For each variation, the branching fraction is recalculated, and the difference with the nominal value is taken as the systematic uncertainty associated with that parameter.
The smoothing parameters  of the non-parametric functions are also varied. The differences in the fit results are included as systematic uncertainties. We add  all uncertainties in quadrature to obtain the overall uncertainty due to PDF parametrization. 
The uncertainties due to errors in the calibration factors used to account for data-MC differences in the signal PDF are evaluated separately but in a similar manner. 
To test the stability of our fitting procedure, we generate and fit a large  ensemble of pseudoexperiments. We find a potential fit bias of $+2.1$\%. We attribute this bias to  neglecting  small correlations among the fitted observables. 
We assign a 1.5\%
systematic uncertainty  due to $\piz$ reconstruction; this is determined from a study of $\tau^-\to\pi^-\piz\nu_{\tau}$ decays~\cite{Ryu:2014vpc}.
The systematic uncertainty due to the track reconstruction efficiency is 0.35\% per track, as determined from a study of partially reconstructed
$D^{*+}\to D^0\pip,\Dz\to\KS\pip\pim$ decays. 
A  0.6\% systematic uncertainty is assigned  due to the particle identification efficiency of the proton-antiproton pair; this is determined from a study of $\Lambda\to p\pim$ decays.  We determine the systematic
uncertainty due to the $C_{\rm NN}$ selection by applying different
$C_{\rm NN}$ criteria and comparing the results with that of the $C_{\rm NN}$
nominal selection. 
The uncertainty due to the estimated fraction of incorrectly reconstructed signal 
events is obtained by varying this fraction by $\pm 50\%$.
The systematic uncertainty due to the total number of $\B\Bb$ pairs is 1.4\%, and the uncertainty due to MC used to evaluate the reconstructed efficiency is 0.4\%.  The total systematic uncertainty is obtained by adding each source in quadrature, as they are assumed to be uncorrelated.
\begin{table}[htb]
\renewcommand{\arraystretch}{1.2}
\caption{\small Systematic uncertainties in $\mathcal{B}(\Bz\to p\bar{p}\pi^0)$. Those listed in the upper section are associated with fitting for the signal yields and are included in the signal significance.}
\label{tab:sys1}
\centering
\begin{tabular}{l|c}
\hline \hline
Source & Uncertainty (\%) \\
\hline
PDF parametrization & $^{+2.9}_{-3.2}$\\
Calibration factor & $11.9$\\
Fit bias & $^{+2.1}_{-0.0}$\\
\hline
$\pi^0$ reconstruction & 1.5 \\
Tracking  & 0.7 \\
Particle identification & $0.6$\\
Choice of $C_{\rm NN}$ &$^{+2.0}_{-1.1}$\\
Incorrectly reconstructed signal events & $^{+1.0}_{-0.8}$\\ 
Number of $B\Bbar$ pairs & 1.4\\
MC statistics & 0.4 \\
\hline
Total & $^{+12.8}_{-12.6}$\\
\hline\hline
\end{tabular}
\end{table}

\begin{figure}[h!t!p!]
\begin{center}
    \includegraphics[width=0.5\textwidth]{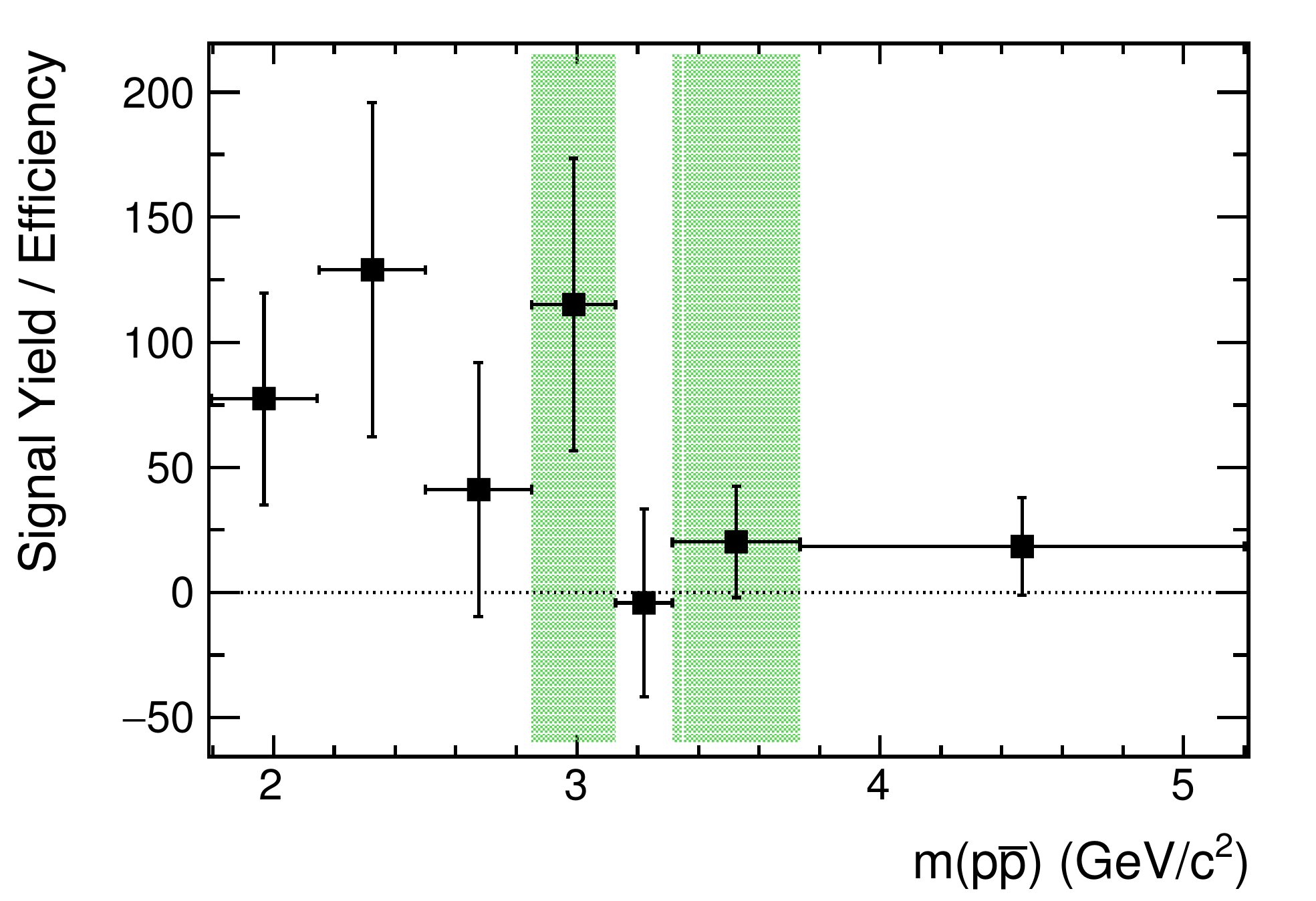}
\end{center}\label{fig:splot_pp}
\vskip -0.5cm
\caption{\small Background-subtracted  and efficiency-corrected distribution of $m(p\bar{p})$.  Points with   error bars are the data and (green) shaded regions represent the area of charmonium veto.}
\end{figure}

\begin{figure}[h!t!p!]
\begin{center}
    \includegraphics[width=0.49\textwidth]{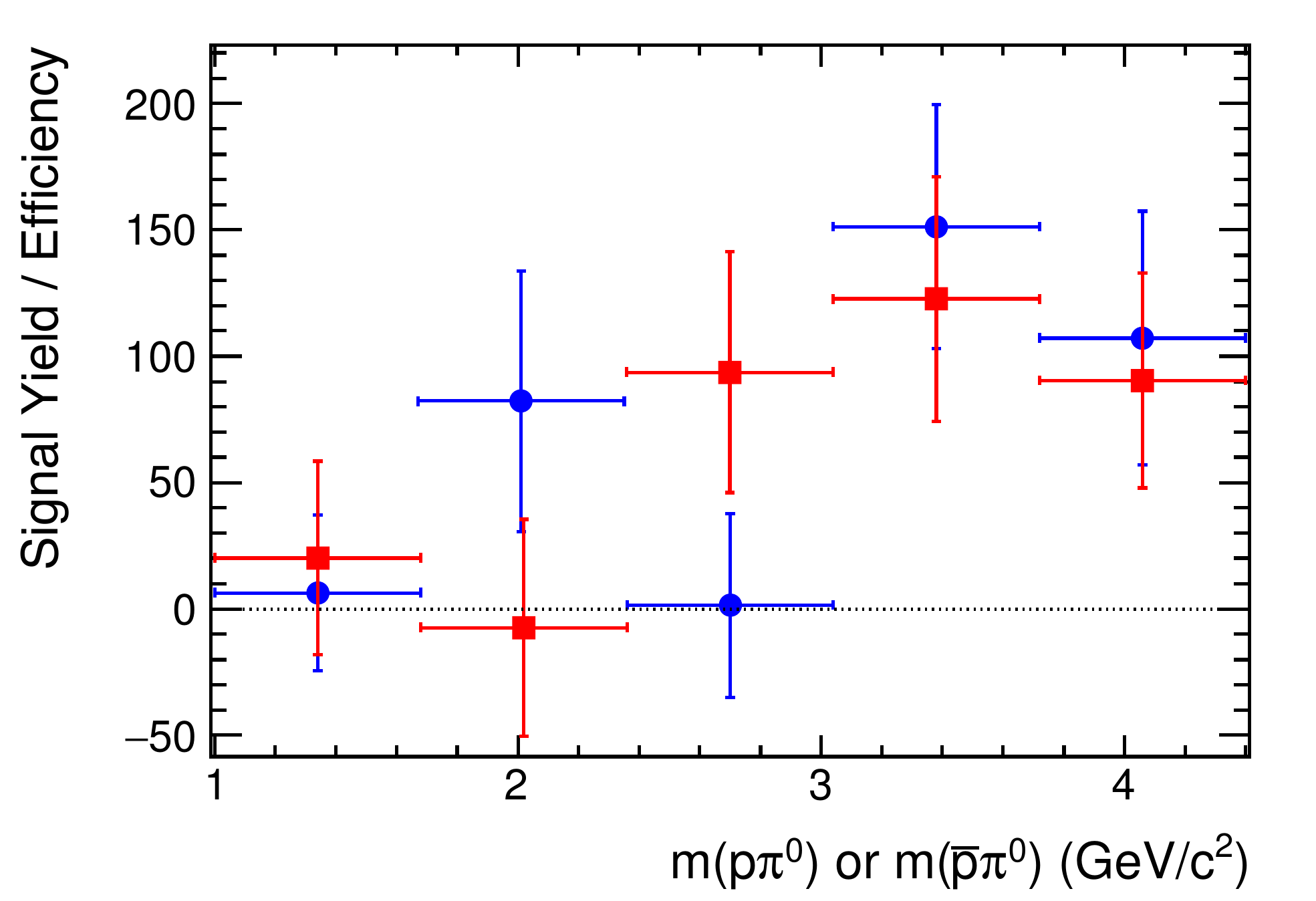}%
\end{center}\label{fig:splot_ppiz}
\vskip -0.5cm
\caption{\small Background-subtracted  and efficiency-corrected distributions of $m(p\piz)$ and $m(\bar{p}\piz)$.  The (blue) circles represent  $m(p\piz)$ and (red) squares the $m(\bar{p}\piz)$. The  charmonium veto is not applied in this plot.}
\end{figure}

Figure~\ref{fig:splot_pp} shows the background-subtracted  and efficiency-corrected  distribution of $m(p\bar{p})$, where  the charmonium veto is removed. 
For the background subtraction, we use the {\it sPlot} technique~\cite{Pivk:2004ty}, with $M_{\rm bc}$, $\Delta E$ and $C'_{\rm NN}$ as the
discriminating variables.
As expected, an enhancement near threshold is visible.
The background-subtracted distributions of $m(p\piz)$ and $m(\bar{p}\piz)$  are shown in Fig.~\ref{fig:splot_ppiz}. 
No obvious structure is seen in these distributions. 

We also search for the intermediate two-body decay $\Bz\to \Delta^+(\to p\piz)\bar{p}$. Events with $m(p\piz)<1.4$~\gevcc are selected for this search. 
No significant signal is observed in this mass range.
We set an upper limit on the branching fraction of $\Bz\to \Delta^+\bar{p}$  at 90\% confidence level (C.L.) using a Bayesian approach. The limit is obtained by integrating the likelihood function from zero to infinity; the value that corresponds to 90\% of this total area is taken as the 90\% C.L. upper limit. We include the systematic uncertainty in the calculation by convolving the likelihood distribution with a Gaussian function whose width is set equal to the total systematic uncertainty of $\mathcal{B}(\Bz\to p\bar{p}\piz)$. As we do not know the flavor of the $\B$ meson at decay, we express our result as a sum of final states containing either a $\Delta^+$ or a $\bar{\Delta}^-$. The result is 
\begin{linenomath}
\begin{equation*}
\mathcal{B}(\Bd\to \Delta^+\bar{p})+\mathcal{B}(\Bd\to\bar{\Delta}^-p)<1.6\times10^{-6}.
\end{equation*}
\end{linenomath}
This is the first such limit and is in agreement with the theoretical predictions~\cite{Chernyak:1990ag, Cheng:2001tr}.

In summary, using the full set of Belle data,  we report a measurement of the branching fraction for $\Bd\to p\bar{p}\piz$ decays. We obtain
$\mathcal{B}(\Bd\to p\bar{p}\pi^0)= (5.0\pm1.8\pm0.6 )\times 10^{-7}$, 
where the first uncertainty is statistical and the second is systematic. The significance of this result is 3.1 standard deviations, and thus this measurement constitutes the first evidence for this decay. 
We also search for the intermediate two-body decays $\Bd\to\Delta^+\bar{p}$ and  $\Bd\to\bar{\Delta}^-p$, and set an upper limit  on the branching fraction, $\mathcal{B}(\Bd\to \Delta^+\bar{p})+\mathcal{B}(\Bd\to\bar{\Delta}^-p)<1.6\times10^{-6}$ at 90\% C.L.

We thank the KEKB group for the excellent operation of the
accelerator; the KEK cryogenics group for the efficient
operation of the solenoid; and the KEK computer group, and the Pacific Northwest National
Laboratory (PNNL) Environmental Molecular Sciences Laboratory (EMSL)
computing group for strong computing support; and the National
Institute of Informatics, and Science Information NETwork 5 (SINET5) for
valuable network support.  We acknowledge support from
the Ministry of Education, Culture, Sports, Science, and
Technology (MEXT) of Japan, the Japan Society for the 
Promotion of Science (JSPS), and the Tau-Lepton Physics 
Research Center of Nagoya University; 
the Australian Research Council including grants
DP180102629, 
DP170102389, 
DP170102204, 
DP150103061, 
FT130100303; 
Austrian Science Fund under Grant No.~P 26794-N20;
the National Natural Science Foundation of China under Contracts
No.~11435013,  
No.~11475187,  
No.~11521505,  
No.~11575017,  
No.~11675166,  
No.~11705209;  
Key Research Program of Frontier Sciences, Chinese Academy of Sciences (CAS), Grant No.~QYZDJ-SSW-SLH011; 
the  CAS Center for Excellence in Particle Physics (CCEPP); 
the Shanghai Pujiang Program under Grant No.~18PJ1401000;  
the Ministry of Education, Youth and Sports of the Czech
Republic under Contract No.~LTT17020;
the Carl Zeiss Foundation, the Deutsche Forschungsgemeinschaft, the
Excellence Cluster Universe, and the VolkswagenStiftung;
the Department of Science and Technology of India; 
the Istituto Nazionale di Fisica Nucleare of Italy; 
National Research Foundation (NRF) of Korea Grants
No.~2015H1A2A1033649, No.~2016R1D1A1B01010135, No.~2016K1A3A7A09005
603, No.~2016R1D1A1B02012900, No.~2018R1A2B3003 643,
No.~2018R1A6A1A06024970, No.~2018R1D1 A1B07047294; Radiation Science Research Institute, Foreign Large-size Research Facility Application Supporting project, the Global Science Experimental Data Hub Center of the Korea Institute of Science and Technology Information and KREONET/GLORIAD;
the Polish Ministry of Science and Higher Education and 
the National Science Center;
the Grant of the Russian Federation Government, Agreement No.~14.W03.31.0026; 
the Slovenian Research Agency;
Ikerbasque, Basque Foundation for Science, Spain;
the Swiss National Science Foundation; 
the Ministry of Education and the Ministry of Science and Technology of Taiwan;
and the United States Department of Energy and the National Science Foundation.


\begin{thebibliography}{99}
\bibitem{charge-conjugate}
Unless stated otherwise, charge-conjugate modes are implicitly included.
\end{thebibliography}
\end{document}